\documentclass[useAMS,usenatbib]{mnras}
\usepackage[T1]{fontenc}
\usepackage{amsmath, amssymb, bm}
\usepackage{physics}
\usepackage[dvipsnames,table]{xcolor}
\usepackage{graphicx}
\usepackage{caption}
\usepackage{subcaption}
\usepackage{stfloats}
\usepackage{booktabs}
\usepackage{tabularx}
\usepackage{multicol}
\usepackage{hyperref}
\usepackage[capitalize]{cleveref}
\usepackage{acro}
\usepackage{xspace}
\usepackage[group-separator={,}]{siunitx}
\usepackage{url}
\usepackage{orcidlink}

\hypersetup{breaklinks=true}

\newcommand{\zCMB}{z_{\rm CMB}}
\newcommand{\zcosmo}{z_{\rm cosmo}}
\newcommand{\zpec}{z_{\rm pec}}
\newcommand{\dlim}{d_{\rm lim}}

\newcommand{\zlim}{z^{\rm lim}}
\newcommand{\dpred}{\bm{d}_{\rm pred}}

\newcommand{\Vpec}{V_{\rm pec}}
\newcommand{\Vext}{\bm{V}_{\rm ext}}

\newcommand{\Mpch}{\ensuremath{\,h^{-1}\,\mathrm{Mpc}}}

\newcommand{\invMpch}{\ensuremath{\,h\,\mathrm{Mpc}^{-1}}}

\newcommand{\kmsec}{\ensuremath{\,\mathrm{km}\,\mathrm{s}^{-1}}}
\newcommand{\kmsecMpc}{\ensuremath{\,\mathrm{km}\,\mathrm{s}^{-1}\,\mathrm{Mpc}^{-1}}}

\newcommand{\Om}{\Omega_{\rm m}}
\newcommand{\PP}{Pantheon\texttt{+}}
\newcommand{\TWOMPP}{2M\texttt{++}}
\newcommand{\Manticore}{\texttt{Manticore-Local}}

\defcitealias{Carrick_2015}{C15}
\defcitealias{VF_olympics}{S25}
\defcitealias{Kenworthy_2022}{K22}
\defcitealias{Riess_2022}{R22}

\DeclareAcronym{CMB}{short = CMB, long  = cosmic microwave background}
\DeclareAcronym{LOS}{short = LOS, long  = line-of-sight}
\DeclareAcronym{BORG}{short = \texttt{BORG}, long  = \textit{Bayesian Origin Reconstruction from Galaxies}}
\DeclareAcronym{SDSS}{short = SDSS, long = Sloan Digital Sky Survey}
\DeclareAcronym{LCDM}{short = $\Lambda$CDM, long  = $\Lambda$-cold dark matter}
\DeclareAcronym{HMC}{short = HMC, long  = Hamiltonian Monte Carlo}
\DeclareAcronym{NUTS}{short = NUTS, long  = No-U-Turn Sampler}
\DeclareAcronym{MCMC}{short = MCMC, long = Markov Chain Monte Carlo}
\DeclareAcronym{CPLR}{short = CPLR, long = Cepheid Period-Luminosity Relation}
\DeclareAcronym{CDF}{short = CDF, long = cumulative distribution function}
\DeclareAcronym{PCS}{short = PCS, long = piecewise-cubic-spline}
\DeclareAcronym{LMC}{short = LMC, long = Large Magellanic Cloud}
\DeclareAcronym{MW}{short = MW, long = Milky Way}
\DeclareAcronym{HST}{short = HST, long = Hubble Space Telescope}
\DeclareAcronym{TRGB}{short = TRGB, long = Tip of the Red Giant Branch}
\DeclareAcronym{SBF}{short = SBF, long = Surface Brightness Fluctuations}
\DeclareAcronym{SN}{short = SN, long  = supernova, short-plural = e, long-plural  = e}

\title[$H_0$ from Cepheids alone]{Two per cent measurement of $H_0$ from Cepheids alone}

\author[R. Stiskalek et al]{Richard Stiskalek$^{1}$\thanks{\href{mailto:richard.stiskalek@physics.ox.ac.uk}{richard.stiskalek@physics.ox.ac.uk}}\orcidlink{0000-0002-0986-314X},
Harry Desmond$^{2}$\thanks{\href{mailto:harry.desmond@port.ac.uk}{harry.desmond@port.ac.uk}}\orcidlink{0000-0003-0685-9791},
Eleni Tsaprazi$^{3}$\orcidlink{0000-0001-5082-4380},
Alan Heavens$^{3}$\orcidlink{0000-0003-1586-2773},\newauthor
Guilhem Lavaux$^{4}$\orcidlink{0000-0003-0143-8891},
Stuart McAlpine$^{5}$\orcidlink{0000-0002-8286-7809} and
Jens Jasche$^{5}$\orcidlink{0000-0002-4677-5843}
\\
$^{1}$Astrophysics, University of Oxford, Denys Wilkinson Building, Keble Road, Oxford, OX1 3RH, UK\\
$^{2}$Institute of Cosmology \& Gravitation, University of Portsmouth, Dennis Sciama Building, Portsmouth, PO1 3FX, UK\\
$^{3}$Imperial Centre for Inference and Cosmology (ICIC) \& Imperial Astrophysics, Department of Physics, Imperial College,\\Blackett Laboratory, Prince Consort Road, London SW7 2AZ, UK.\\
$^{4}$CNRS \& Sorbonne Universit\'e, Institut d'Astrophysique de Paris (IAP), UMR 7095, 98 bis bd Arago, F-75014 Paris, France\\
$^{5}$The Oskar Klein Centre, Department of Physics, Stockholm University, Albanova University Center, 106 91 Stockholm, Sweden
}

\date{Accepted XXX. Received YYY; in original form ZZZ}
\pubyear{2025}

\begin{document}\label{firstpage}
\pagerange{\pageref{firstpage}--\pageref{lastpage}}
\maketitle

\begin{abstract}
One of the most pressing problems in current cosmology is the cause of the Hubble tension. We revisit a \emph{two-rung distance ladder}, composed only of Cepheid periods and magnitudes, anchor distances in the Milky Way, Large Magellanic Cloud, NGC\,4258, and host galaxy redshifts. We adopt the SH0ES data for the most up-to-date and carefully vetted measurements, where the Cepheid hosts were selected to harbour also Type Ia supernovae. We introduce two important improvements: a rigorous selection modelling and a state-of-the-art density and peculiar velocity model using \Manticore, based on the \emph{Bayesian Origin Reconstruction from Galaxies} (\texttt{BORG}) algorithm. We infer $H_0 = 71.1 \pm 1.4\kmsecMpc$, assuming the Cepheid host sample was selected by supernova magnitudes.
However, the actual selection criteria are not clear, and other assumptions can increase $H_0$ by up to one statistical standard deviation.
The posterior has a lower central value and a 41 per cent smaller uncertainty than a previous study using the same distance-ladder data. This result is lower than the supernova-based SH0ES inferred value of $H_0 = 73.2 \pm 0.9\kmsecMpc$ at about $1.3\sigma$, and is in $2.8\sigma$ tension with the latest cosmic microwave background results in the standard cosmological model. These results demonstrate that a measurement of $H_0$ of sufficient precision to weigh in on the Hubble tension is achievable using second-rung data alone, underscoring the importance of robust and accurate statistical and velocity-field modelling.
\end{abstract}



\section{Introduction}\label{sec:intro}

The Hubble tension is potentially the most serious challenge faced by the concordance model of cosmology, \ac{LCDM}. This is a claimed $5.8\sigma$ discrepancy~\citep{Breuval_2024} between the present expansion rate of the Universe---the Hubble parameter $H_0$---inferred from the \ac{CMB} by the \textit{Planck} satellite~\citep{Planck_2020_cosmo,Tristram_2024} versus the local distance ladder constructed from Cepheids and Type Ia \acp{SN} in the \textit{Supernovae and $H_0$ for the Equation of State of dark energy} programme (SH0ES;~\citealt{Riess_2022}). A vast array of models have been proposed in response to this, ranging from the addition of a dark energy-like component before recombination to modified gravity in Cepheid stars, yet none is able to resolve the tension while retaining consistency with other observations (for reviews see~\citealt{hubble_realm,H0_olympics,cosmoverse}).
\begingroup
\renewcommand{\thefootnote}{}
\footnotetext{This updated version of the manuscript follows an erratum correcting the treatment of the 3D source-position prior and the associated selection function. The main conclusions are unchanged. The fiducial \Manticore\ supernova-magnitude-selected result shifts down to $H_0 = 71.1 \pm 1.4\kmsecMpc$, and the redshift-selected result shifts to $72.5 \pm 1.4\kmsecMpc$, corresponding to ${\sim}0.5\sigma$ shifts with respect to the previously published values. This version also adopts the 80 COLA \Manticore\ realisations in place of the 30 \texttt{SWIFT}-based posterior $N$-body resimulations used previously.}
\endgroup

A persistent concern is that the Hubble tension may arise not from new physics, but rather from unknown systematics or other modelling deficiencies in the $H_0$ inference pipelines. This has led to a battery of cross-checks on both the \ac{CMB} and distance ladder sides, which has not however yielded any definitive conclusion. Alternative \ac{CMB} data and analysis methods corroborate the \textit{Planck} measurement to high precision~\citep{WMAP,Planck_intermediate,ACT_H0,SPT_H0,Efs_planck,Calabrese_2025,Camphuis_2025}, a range of systematics have been investigated and found not to impact the inferred $H_0$ significantly~\citep{Riess_2022,Riess_2022B,Riess_2023,Bhardwaj_2023,Riess_2024,Najeira_2025,Carreres_2025,Tsaprazi_2025}, and multiple alternative reconstructions of the local distance ladder---and other low-redshift methods---prefer a higher $H_0$, albeit with lower precision than SH0ES (e.g.~\citealt{Burns,megamaser, Schombert,Blakeslee_2021,Dhawan,Jaegar,Vogl_2024,Jensen_2025}). These studies have consolidated the belief that the Hubble tension is real and demands an explanation invoking new physics rather than deficiencies in the data analysis.

A key approach for assessing the robustness of the SH0ES inference---and that of the local distance ladder more generally---is to drop or swap out various of the star types used. In particular the \acp{SN}, which in SH0ES extend the Hubble diagram from $z=0.0233$ to $z=0.15$, are sometimes considered a weak link: they require standardisation based on light curve and environmental properties, complex dust corrections, absolute-magnitude calibration from lower-rung distance data and complex simulation-based selection modelling (e.g.~\citealt{SN1,SN2,SN3,SN4,Efs_SNe,Wojtak_2025}). This raises the question of whether $H_0$ may be inferred (albeit to lower precision) without them, which was in fact the original method of~\cite{Hubble}, and also employed by~\cite{Willick_2001}, who measured $H_0$ from geometry- and Cepheid-based distances---just as we do here. A discrepancy could then imply systematics in the full SH0ES analysis; as this is the only $H_0$ analysis that exhibits a statistically significant discrepancy with \textit{Planck}, this has the potential to pull the rug out from under the Hubble tension itself.

This approach was taken earlier by~\citet{Kenworthy_2022} (hereafter~\citetalias{Kenworthy_2022}). This study took 35 galaxies with both Cepheid and \ac{SN} measurements from the SH0ES project. In SH0ES the Cepheids are used merely to calibrate the \ac{SN} absolute magnitude (i.e.\,the redshifts of the hosts are unused); here, by contrast, the \acp{SN} are discarded and $H_0$ is inferred from the Cepheid distances and host galaxy redshifts. The \acl{CPLR} (\ac{CPLR};~\citealt{Leavitt_1912}) is calibrated from the first, geometric rung of the distance ladder, namely parallax using \textit{Gaia} within the \ac{MW}~\citep{Riess_2018, Riess_2021}, detached eclipsing binaries within the \ac{LMC}~\citep{Pietrzynski_2019} and the water maser in NGC\,4258~\citep{Reid_2019}. At these distances, peculiar velocities are a significant perturbation to the cosmological redshifts and must be modelled.~\citetalias{Kenworthy_2022} develop and deploy models for these peculiar velocities, as well as possible sample selection effects, to find an overall result $H_0 = 72.9^{+ 2.4}_{-2.2}\kmsecMpc$. This is in agreement with the SH0ES constraint and in mild, 2.3$\sigma$ tension with \textit{Planck}, leading the authors to conclude that the \acp{SN} in SH0ES are not appreciably altering the best-fit $H_0$, and therefore are not responsible for the Hubble tension.

The present study is a reanalysis of that data and a reassessment of that conclusion. We construct a Bayesian forward model of the distance ladder for inferring $H_0$ (and a slew of nuisance parameters) from the Cepheid properties, geometric anchor data and host galaxy redshifts. We present a principled Bayesian framework to account for sample selection and demonstrate its impact on the inferred $H_0$. In addition, we employ state-of-the-art peculiar velocity modelling with the \Manticore\ reconstruction. In comparison with~\citetalias{Kenworthy_2022}, we find a lower central value of $H_0$, and a much reduced error bar. To understand the differences, we highlight several crucial improvements over their analysis.

The structure of the paper is as follows. \Cref{sec:data} describes the SH0ES data that we use, which is identical to that of~\citetalias{Kenworthy_2022}. \Cref{sec:methodology} presents our method: the inference framework (\cref{sec:forward_model}), peculiar velocity models (\cref{sec:PV_models}), and selection function modelling (\cref{sec:method_selection_function}). \Cref{sec:results} details our results: we set the stage with a distance-only analysis not involving redshifts (\cref{sec:Cepheid_only_distances}), before moving onto the $H_0$ inference and its robustness to peculiar velocity modelling and selection effects (\cref{sec:Cepheid_only_H0}). \Cref{sec:discussion_selection} discusses the importance of selection function modelling, \cref{sec:kenworthy_comparison} provides a detailed comparison with the method of~\citetalias{Kenworthy_2022},~\cref{sec:ramifications} discusses the broader implications of our results and suggests further work, and~\cref{sec:conclusion} concludes. Appendix~\ref{sec:host_table} lists the properties of the Cepheid host galaxies used in our analysis, Appendix~\ref{sec:mock_data_bias_tests} details the mock tests that we develop to demonstrate that our model is unbiased, Appendix~\ref{sec:universe_reconstructions} describes the \Manticore\ and~\cite{Carrick_2015} peculiar velocity reconstructions and Appendix~\ref{sec:LCDM_covariance} summarises the calculation of the \ac{LCDM} peculiar velocity covariance.

\vspace{1em}
\noindent All logarithms are base-10 unless otherwise stated. We use the notation $\mathcal{N}(x; \mu, \sigma)$ to denote a one-dimensional normal distribution with mean $\mu$ and standard deviation $\sigma$, evaluated at $x$; in higher dimensions $\mu$ is a vector and $\sigma$ is replaced by a covariance matrix. We define $h \equiv H_0\ / \left(100\kmsecMpc\right)$.


\section{Data}\label{sec:data}

We use data from the SH0ES programme~\citep{Riess_2022}, which aims to measure the Hubble constant $H_0$ using the local distance ladder. The catalogue combines observations of Cepheids in the \ac{MW} and nearby galaxies, including hosts with both Cepheids and Type Ia \acp{SN}.

The \ac{CPLR} is anchored by \ac{MW} parallaxes~\citep{Riess_2018, Riess_2021}, detached eclipsing binaries in the \ac{LMC}~\citep{Pietrzynski_2019}, and megamasers in NGC\,4258~\citep{Reid_2019}.~\citet{Reid_2019} measure a distance modulus to NGC\,4258 of $29.398 \pm 0.032$ mag, while~\citet{Pietrzynski_2019} determine a distance modulus to the \ac{LMC} of $18.477 \pm 0.026$ mag; both are adopted by SH0ES as geometric anchors for the Cepheid calibration.\footnote{Recently, the Small Magellanic Cloud (SMC) has also been used as a geometric anchor by~\cite{Breuval_2024}.} SH0ES also incorporates 55 Cepheids in M31 observed with the \ac{HST} using the same three-filter system, as presented by~\citet{Li_2021}. This sample provides precise measurements that constrain the slope of the period--luminosity relation.

Beyond the \ac{MW}, \ac{LMC}, M31, and NGC\,4258, the SH0ES sample includes 35 galaxies hosting both Cepheids and Type Ia \acp{SN}. An additional 254 Type Ia \acp{SN}, observed in galaxies without Cepheids, extend the measurement into the Hubble flow, with their calibration tied to the Cepheid-\ac{SN} hosts. However, in this work we do not use any SH0ES \ac{SN} data, but instead focus on a two-rung distance ladder involving only geometric anchors and Cepheids. This has the crucial advantage of affording a \ac{SN}-independent determination of $H_0$, helping to check for potential systematics in the \ac{SN} analysis. We now provide a brief overview of the geometric anchor and Cepheid distance ladder in SH0ES.

The $j$\textsuperscript{th} Cepheid in the $i$\textsuperscript{th} host has apparent magnitude
\begin{equation}\label{eq:cepheid_magnitude}
    m_{W,ij} = \mu_i + M_W + b_W \log \left(P_{ij} / \hat{P}\right) + Z_W [\mathrm{O/H}]_{ij},
\end{equation}
where $\mu_i$ is the host distance modulus, $M_W$ is the fiducial absolute magnitude of a Cepheid with a period of 10 days and solar metallicity (zero-point), $b_W$ is the period--luminosity slope, $P_{ij}$ is the Cepheid period, $\hat{P} = 10~\mathrm{days}$, $Z_W$ is the coefficient of the metallicity correction, and $[\mathrm{O/H}]_{ij}$ is the metallicity at the Cepheid's galactocentric position.

To account for systematic uncertainty in the metallicity, SH0ES propagates the difference between the mean of nine calibrations of strong-line abundance measurements and the scale of~\citet{Pettini_2004}, which matches well with direct extragalactic stellar abundances~\citep{Bresolin_2016}.

This dependence is encoded in the covariance matrix of the apparent magnitudes, which depends on $Z_W$, whose fiducial value in~\citet{Riess_2022} is determined from the joint optimisation of the geometric anchor, Cepheid, and \ac{SN} components of the SH0ES distance ladder. We adopt the same approach and neglect the dependence of the covariance on $Z_W$ during inference, assuming the same covariance matrix with a fiducial $Z_W$ as~\citet{Riess_2022}.

An additional contribution to the Cepheid magnitude covariance arises from uncertainties in the sky background estimation due to crowding. This is quantified by injecting artificial stars near each Cepheid at expected magnitudes from trial period--luminosity fits. The discrepancy between input and recovered magnitudes informs a correction to the photometry and yields a background-induced covariance term for Cepheids in the same host galaxy. The resulting uncertainty, which reflects correlated photometric biases among Cepheids in the same host, contributes an error floor of $0.03$ to $0.06$~mag per host. We denote the final Cepheid magnitude covariance matrix as $\mathbf{\Sigma}_{\rm Ceph}$, which combines the contributions from the systematic uncertainties in the metallicity scale, the background-induced photometric biases, and the intrinsic Cepheid uncertainties. For more details, see Section~2.1 of~\citet{Riess_2022}.

The SH0ES analysis incorporates an external calibration of the fiducial Cepheid absolute magnitude using \ac{MW} Cepheids with trigonometric parallax distances. Two samples are used: eight Cepheids with high-precision \ac{HST} spatial-scan parallaxes~\citep{Riess_2018} and 75 Cepheids with \textit{Gaia} EDR3 parallaxes~\citep{Riess_2021}, both with fluxes measured on the same \ac{HST} photometric system and with direct spectroscopic metallicities. These samples provide independent constraints on $M_W$, denoted $M_{W}^{\rm HST} = -5.804 \pm 0.082$ and $M_{W}^{\rm Gaia} = -5.903 \pm 0.025$, derived from \ac{HST} and \textit{Gaia} parallaxes, respectively. In addition to the \ac{MW} calibration, SH0ES uses Cepheid observations in the \ac{LMC} to further constrain the luminosity scale. A systematic offset $\Delta_{\rm ZP}$ with uncertainty $\sigma_{\mathrm{grnd}} = 0.10~\mathrm{mag}$ is introduced to account for differences between ground- and space-based photometry in the \ac{LMC} calibration. The anchor constraints are summarised in~\cref{tab:external_constraints}.

Unlike the SH0ES analysis, which does not use the redshifts of Cepheid host galaxies but instead infers $H_0$ from more distant \acp{SN}~\citep{Riess_2022}, we infer $H_0$ directly from the redshifts of the Cepheid hosts, without relying on \acp{SN}. Specifically, we use the observed redshifts of the 37 galaxies that host both Cepheids and Type Ia \acp{SN}, excluding the geometric anchors (\ac{LMC}, M31, and NGC\,4258). These redshifts, converted to the \ac{CMB} frame, are taken from the~\PP\ sample, which includes these galaxies as part of its \ac{SN} compilation~\citep{Brout_2022}. Similarly to~\citetalias{Kenworthy_2022}, we exclude the two most distant host galaxies out of the 37, as they were targeted with a different \ac{HST} programme and are therefore subject to a distinct selection function. Our final sample therefore comprises 35 galaxies, whose coordinates are also listed in Appendix~\ref{sec:host_table}.

All Cepheid host galaxies are confined to within approximately 40 Mpc (or redshift less than 0.011), indicating the presence of some selection effects in constructing the sample.
This selection does not arise from incompleteness in \ac{SN} detection at these distances. Rather, the selection reflects the fact that only a (random) subset of detected \acp{SN} with $m_{\rm SN} \lesssim 14$~mag or $cz \lesssim 3300\kmsec$ were selected for \ac{HST} Cepheid follow-up observations.
We verify that it is a random subset by comparing the \ac{SN} magnitudes and redshifts of the Cepheid host galaxies to the full \PP sample.
It is not clear whether this selection is in the \ac{SN} apparent magnitude, Cepheid apparent magnitude, the host galaxy redshift or their combination, given that Cepheid observations were assembled over many years from independent proposals without a unified selection strategy~\citepalias{Kenworthy_2022}. We discuss our approach to modelling the selection function, which differs from that of~\citetalias{Kenworthy_2022}, in~\cref{sec:method_selection_function}. To model selection in \ac{SN} apparent magnitude, we use the observed apparent magnitudes of the \acp{SN} in the 35 host galaxies, adopting the bias-corrected apparent magnitudes from the SH0ES sample together with the corresponding covariance matrix $\mathbf{\Sigma}_{\rm SN}$.

\begin{figure*}
    \centering
    \includegraphics[width=\textwidth]{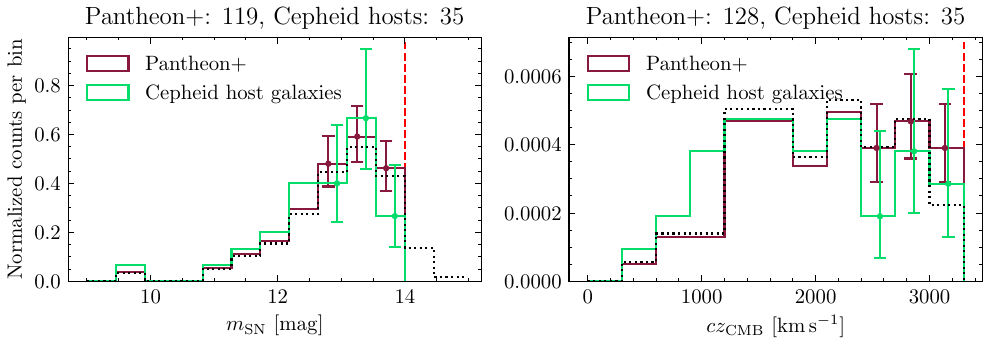}
    \caption{Distribution of \ac{SN} apparent magnitudes for the 35 Cepheid host galaxies and their observed host redshifts either below a limiting \ac{SN} magnitude of 14~mag or host observed redshift of $3300\kmsec$ (left and right panels, respectively). Following a Kolmogorov–Smirnov test, the two samples are mutually consistent with $p = 0.94$ and $p=0.60$. Assuming that the \PP\ sample is complete within ${\sim} 40\,\mathrm{Mpc}$, this implies random selection of Cepheid hosts below a limiting \ac{SN} apparent magnitude of 14 and a host observed redshift of $3300\kmsec$. The error bars are $1\sigma$ Poisson counting errors. In addition, the dotted black lines show the magnitude distribution of \PP\ hosts with observed redshifts below $3300~\kmsec$ in the left panel, and the redshift distribution of hosts with magnitudes below 14 in the right panel. Within \PP, a truncation in magnitude produces a sharply truncated redshift distribution, whereas a truncation in redshift yields a small tail towards higher magnitudes. We therefore regard \ac{SN} magnitude selection as marginally more likely and adopt it as our fiducial scenario.}
    \label{fig:SH0ES_sample}
\end{figure*}

In~\cref{fig:SH0ES_sample} we show the distribution of the \ac{SN} apparent magnitudes for the 35 host galaxies and the distribution of their host observed redshifts. The left panel compares the apparent magnitudes of the SH0ES \acp{SN} to those of the full \PP\ sample. We assume that \PP\ is complete at such small redshifts ($z < 0.011$) down to $m_\mathrm{SN}=14$ mag. The distribution of the 35 Cepheid hosts' \ac{SN} magnitudes, which are below $m_{\rm SN} = 14$~mag, is consistent with \PP\ ($p = 0.94$ from a Kolmogorov–Smirnov test), which implies that it is plausible that the Cepheid host sample has been selected with a hard limiting estimated magnitude of $m_\mathrm{SN}=14$~mag. The same trend can be seen in the right-hand side panel, where we compare the redshift distributions between the \PP\ and Cepheid host samples (mutually consistent with $p = 0.60$).

Although the real selection criteria may be complicated and not universal, further support for modelling the selection on the basis of apparent SN magnitude comes from the distributions of magnitude and redshift. Had the sample been selected on either the basis of magnitudes or redshifts but not both, a tail would be visible in the other distribution. However, as shown in~\cref{fig:SH0ES_sample}, the subset of \PP\ with \ac{SN} magnitudes below 14 leads to a sharp truncation in the observed redshift distribution at $3300\kmsec$, whereas the subset of \PP\ hosts with redshift below $3300\kmsec$ shows a small tail in the \ac{SN} apparent magnitude above 14. We therefore conclude on the basis of current evidence that \ac{SN} selection is the most likely model.
It was also a primary model considered by~\citetalias{Kenworthy_2022}. Both the magnitude and redshift distributions show a small deficit relative to \PP\ near the limiting values. However, these deficiencies are not statistically significant if accounting for the expected Poisson uncertainties. Nevertheless, we account for non-trivial selection near the edge by modelling a smooth detection probability function rather than assuming a sharp selection threshold.

\begin{table*}
    \centering
    \begin{tabular}{p{6cm}p{5cm}p{4cm}}
    \textit{Constraint} & \textit{Value} & \textit{Reference} \\
    \toprule
    HST Milky Way Cepheid zero-point  & $-5.804 \pm 0.082$~mag & \citet{Riess_2018} \\
    \textit{Gaia} Milky Way Cepheid zero-point  & $-5.903 \pm 0.025$~mag & \citet{Riess_2021} \\
    LMC distance modulus & $18.477 \pm 0.026$~mag & \citet{Pietrzynski_2019} \\
    NGC\,4258 distance modulus & $29.398 \pm 0.032$~mag & \citet{Reid_2019} \\
    \bottomrule
    \end{tabular}
    \caption{External observational constraints entering the Bayesian hierarchical model. The \ac{HST} and \textit{Gaia} \ac{MW} calibrations constrain the Cepheid absolute magnitude zero-point, while the geometric anchors constrain the distances to the anchor galaxies \ac{LMC} and NGC\,4258. We use these measurements as they have been reported by~\protect\citet{Riess_2022}.}
    \label{tab:external_constraints}
\end{table*}

\section{Methodology}\label{sec:methodology}

\subsection{Bayesian forward modelling framework}\label{sec:forward_model}

\begin{figure*}
    \centering
   \includegraphics[width=\textwidth]{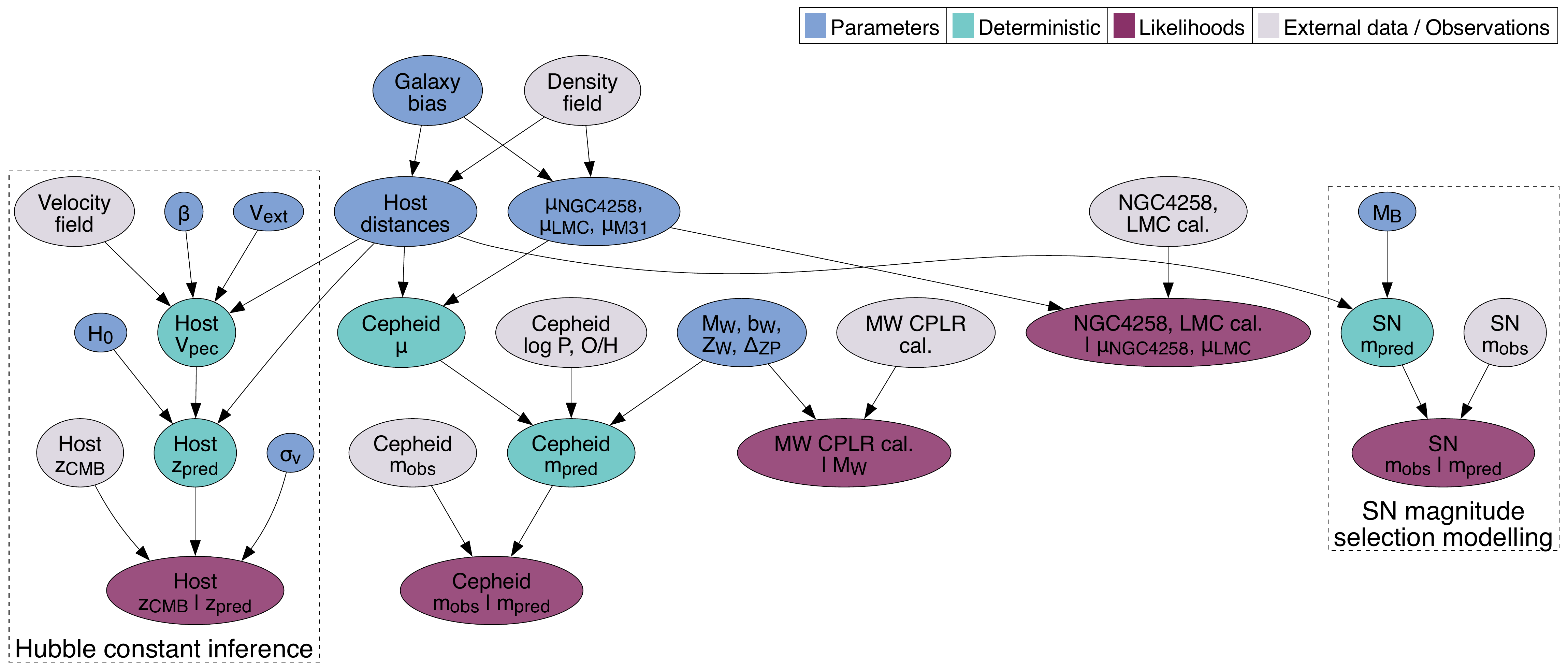}
    \caption{Directed acyclic graph of the probabilistic model used to forward model the Cepheid magnitudes and the host galaxy redshifts. The left-hand dashed black box delineates the portion of the model used to constrain $H_0$. Excluding it yields the analysis in~\cref{sec:Cepheid_only_distances}, where we infer Cepheid host galaxy distances using only the \ac{CPLR} and the two geometric anchors (\ac{LMC} and NGC\,4258). See~\cref{sec:method_selection_function} for details on our modelling of sample selection. The dashed black box on the right marks the \ac{SN} magnitude selection, for which we sample $M_B$ and forward-model the supernova apparent magnitudes to model the selection. When assuming redshift selection for the host sample, this step is omitted.}
    \label{fig:CH0_DAG}
\end{figure*}

We now describe our Bayesian hierarchical model. Our approach differs from that of~\citetalias{Kenworthy_2022} in that, rather than using pre-computed distances from the fiducial SH0ES analysis~\citep{Riess_2022} to predict the observed redshifts of the Cepheid host galaxies in a two-step inference, we forward-model the Cepheid observables directly in a self-contained single-step inference. The data consists of Cepheid magnitudes, periods, and metallicities, and the estimated redshifts of their host galaxies. We also include as observational inputs the \ac{MW} calibration of the Cepheid absolute magnitude, and the geometric distance moduli of the \ac{LMC} and NGC\,4258. Under the assumption of the \ac{SN} magnitude selection of the host galaxy sample, we also include these magnitudes as inputs to our model, as detailed in~\cref{sec:SN_mag_redshift_selection}. The model parameters for this two-rung distance ladder are the Cepheid absolute magnitude $M_W$, the period-luminosity slope $b_W$, the metallicity dependence $Z_W$, and the Hubble constant $H_0$ as well as the standardised \ac{SN} magnitude $M_B$ in the case that we model \ac{SN} selection. Each host galaxy is assigned a latent distance modulus $\mu_i$, providing another 35 parameters to infer. Additional free parameters relate to the peculiar velocity field, for which we consider several models as described in~\cref{sec:PV_models}. In~\cref{sec:method_selection_function}, we describe our approach to modelling selection effects in either \ac{SN} magnitude, Cepheid magnitude or host galaxy redshift. We illustrate the model with a directed acyclic graph in~\cref{fig:CH0_DAG} and summarise the prior distributions of the model's free parameters in~\cref{tab:free_parameters}.

We sample $M_W$, $b_W$, and $Z_W$ from broad uniform priors. Similarly, $H_0$ is sampled uniformly from $10$ to $100\kmsecMpc$. The \ac{MW} Cepheid calibration is imposed through two Gaussian likelihood terms, corresponding to the \ac{HST} and \textit{Gaia} geometric calibration of the fiducial absolute magnitude:
\begin{align}
    \mathcal{L}(M_{W}^{\rm HST} \mid M_W) &= \mathcal{N}(M_{W}^{\rm HST}; M_W, \sigma_{\mathrm{HST}}),\\
    \mathcal{L}(M_{W}^{\rm Gaia} \mid M_W) &= \mathcal{N}(M_{W}^{\rm Gaia}; M_W, \sigma_{\mathrm{Gaia}}),
\end{align}
where $\sigma_{\mathrm{HST}}$ and $\sigma_{\mathrm{Gaia}}$ are the uncertainties associated with the \ac{HST} and \textit{Gaia} calibrations, respectively. We impose priors on the distance moduli $\mu_i$, which describe the distances to the Cepheid host galaxies, and are related to luminosity distance $D_{\mathrm{L},i}$ of the $i$\textsuperscript{th} host as
\begin{equation}
    \mu_i = 5 \log \frac{D_{\mathrm{L},i}}{\mathrm{Mpc}} + 25,
\end{equation}
which is related to comoving distance $r$ as $D_L = (1 + \zcosmo) r$, where $\zcosmo$ is the cosmological redshift. For distances, we choose a prior that is uniform in volume: along a given \ac{LOS}, the probability of finding a galaxy increases as the square of the distance due to the volume element. Sampling the distance moduli $\mu_i$ implies the prior
\begin{equation}
    p(\mu_i) = p(r_i) \left|\dv{r}{\mu}\right|_{\mu_i},
\end{equation}
where $r_i$ is the physical distance, $p(r_i) \propto r_i^2$ and $\left|\dd r/\dd \mu \right|$ is the Jacobian for the transformation from distance to distance modulus.
This prior is important, as we discuss further in~\cref{sec:Cepheid_only_distances}.
More generally, the 3D position prior is $\pi(\bm{r}) \propto n_g(\bm{r})$, where $n_g(\bm{r})$ is the source number density, which we parametrise as a function of the underlying density field.
For the \ac{LMC} and NGC\,4258, we treat the geometric distance calibrations as additional observational constraints with Gaussian likelihoods:
\begin{equation}
    \mathcal{L}(\tilde{\mu}_{k} \mid \mu_{k}) = \mathcal{N}(\tilde{\mu}_{k}; \mu_{k}, \sigma_{k}),
\end{equation}
where $k$ represents either \ac{LMC} or NGC\,4258, $\mu_k$ is the sampled distance modulus, $\tilde{\mu}_k$ is the reported measurement from either~\cite{Pietrzynski_2019} or~\cite{Reid_2019}, respectively, and $\sigma_k$ is the corresponding measurement uncertainty. The Cepheid apparent magnitudes are predicted according to~\cref{eq:cepheid_magnitude}, leading to the likelihood
\begin{equation}
    \mathcal{L}(\bm{m}_W \mid \bm{m}_{W}^{\rm pred}) = \mathcal{N}(\bm{m}_W; \bm{m}_{W}^{\rm pred},\, \mathbf{\Sigma}_{\mathrm{Cep}}),
\end{equation}
where the bold font denotes a vector over all $3130$ Cepheids.

Up to this point, we have used only the geometric anchors and the \ac{CPLR} relation to constrain the physical distances to the Cepheid host galaxies. Constraining $H_0$ requires also including redshift information. The predicted host redshifts, boosted into the \ac{CMB} frame, are given by
\begin{equation}\label{eq:zCMB_pred}
    1 + z_{\rm CMB}^{\rm pred} = (1 + \zcosmo)(1 + \zpec),
\end{equation}
where $\zcosmo$ is the cosmological redshift derived from the host distance modulus. We compute the cosmological redshift assuming a flat \ac{LCDM} cosmology with $\Om = 0.3$ and the sampled value of $H_0$ (though we note that the dependence on $\Om$ is negligible because of the small redshift range of our sample). $\zpec = \Vpec / c$ is the contribution from the \ac{LOS} peculiar velocity $\Vpec$. Assuming Gaussian uncertainty covariance $\mathbf{\Sigma}_{\rm cz}$ of the observed redshifts produces the likelihood term
\begin{equation}\label{eq:zCMB_likelihood}
    \mathcal{L}(\bm{z}_{\rm CMB} \mid \bm{z}_{\rm CMB}^{\rm pred}) = \mathcal{N}(c\bm{z}_{\rm CMB};\, c\bm{z}_{\rm CMB}^{\rm pred},\,\mathbf{\Sigma}_{cz}).
\end{equation}

\vspace{1em}
\noindent To sample the posterior distribution we use the \texttt{numpyro}\footnote{\url{https://num.pyro.ai/en/latest/}} package~\citep{Phan_2019, Bingham_2019}, specifically the \acl{NUTS} method of \acl{HMC} sampling~\citep{Hoffman_2011}. We run twelve independent chains of \num{6000} samples each, discarding the first \num{1000} as burn-in. Convergence is ensured by requiring the Gelman–Rubin statistic $\hat{R}-1 \leq 0.01$ for all parameters.

\subsection{Peculiar velocity modelling}\label{sec:PV_models}

\begin{figure}
    \centering
    \includegraphics[width=\columnwidth]{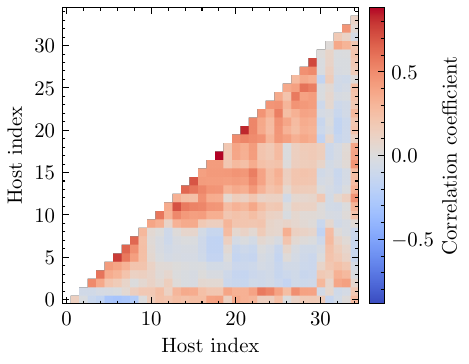}
    \caption{The expected peculiar velocity correlation coefficients computed from the \ac{LCDM} peculiar velocity covariance matrix (see~\cref{sec:LCDM_covariance}) for the 35 Cepheid host galaxies. A large fraction of the peculiar velocities are strongly correlated, reducing the effective sample size and highlighting the need for a local Universe reconstruction such as~\protect\cite{Carrick_2015}.}
    \label{fig:host_covariance}
\end{figure}

Since the Cepheid host galaxies are relatively nearby (the great majority have $cz<3000\kmsec$), peculiar velocities can be a significant contributor to their redshifts. It is therefore important that they are modelled reliably, and that the systematic uncertainties they contribute are reliably assessed. We adopt a series of models to achieve this.
\begin{enumerate}
    \item Our baseline, least realistic model assumes no coherent flows so that $\zpec = 0$ on average. This makes the host galaxy redshifts independent so that the covariance matrix $\mathbf{\Sigma}_{\rm cz}$ is diagonal. We assume that the recession velocity of a given galaxy is Gaussian distributed with a width $\sigma_v$, a free parameter to be inferred. The velocity variances are therefore $\sigma_v^2 + \sigma_{cz}^2$, where $\sigma_{cz}$ is the redshift measurement uncertainty (typically subdominant since $\sigma_v \approx 250\kmsec$). We sample $\sigma_v$ from a reference (scale-invariant) prior, such that
    \begin{equation}
        \pi(\sigma_v) \propto 1 / \sigma_v.
    \end{equation}
    \item This model retains the assumption of a diagonal covariance between predicted and observed redshifts, but explicitly accounts for coherent bulk velocity by modelling the local velocity field as a constant vector $\Vext$. We assign a uniform prior on the magnitude of $\Vext$ and a uniform prior on its direction over the sky. The \ac{LOS} peculiar velocity of the $i$\textsuperscript{th} host is then
    \begin{equation}
        V_{\mathrm{pec},i} = \Vext \cdot \hat{\bm{r}}_i,
    \end{equation}
    where $\hat{\bm{r}}_i$ is the \ac{LOS} unit vector to the $i$\textsuperscript{th} galaxy. We again include $\sigma_v$ sampled from a reference prior.
    \item Rather than explicitly modelling the flow, we use the \ac{LCDM} peculiar velocity covariance matrix $\mathbf{\Sigma}_{\Lambda\mathrm{CDM}}$, thereby marginalising over all possible \ac{LCDM} realisations of the local velocity field. The total velocity covariance is then
    \begin{equation}
        \mathbf{\Sigma}_{cz} = \mathbf{\Sigma}_{\Lambda\mathrm{CDM}} + \mathbf{I} \: (\sigma_v^2 + \sigma_{cz}^2),
    \end{equation}
    where $\sigma_v$ captures residual small-scale velocity dispersion and is sampled from the same reference prior as in previous models and $\mathbf{I}$ is the unit matrix. The construction of $\mathbf{\Sigma}_{\Lambda\mathrm{CDM}}$ is discussed later in this section. We show in~\cref{fig:host_covariance} the correlation coefficients between the peculiar velocities of the 35 Cepheid host galaxies. This is the most conservative model.
    \item Our fourth model extends the previous one by introducing a global scaling parameter $A$ that multiplies $\mathbf{\Sigma}_{\Lambda\mathrm{CDM}}$, following \citet{Huterer_2015}. This is designed to approximate the effect of deviations from the fiducial cosmology used to compute the covariance matrix. By default, we adopt a truncated Gaussian prior on $A$, bounded below at zero, with mean unity and standard deviation 0.5. We also test a uniform prior on $A$ over the range $[0, 5]$, verifying that this choice does not appreciably affect the results. The limiting case $A = 1$ recovers the unscaled \ac{LCDM} model, while $A = 0$ reduces to the $\sigma_v$-only model with diagonal covariance.
    \item Our fifth model uses the reconstruction of~\citet{Carrick_2015} (hereafter~\citetalias{Carrick_2015}), which provides a single realisation of both the density and velocity fields of the local Universe as a function of sky position and radial distance in $h^{-1}\,\mathrm{Mpc}$. The peculiar velocity of the $i$\textsuperscript{th} host is obtained by evaluating the reconstructed velocity field at its sampled position. In this case we also extract the local density from the same reconstruction to model inhomogeneous Malmquist bias. The reconstructed peculiar velocities are scaled by a factor $\beta$, treated as a free parameter with either a uniform prior or a Gaussian prior based on the measurement $\beta = 0.43 \pm 0.02$ from~\citetalias{Carrick_2015}. It is defined as $\beta \equiv f / b_1$, where $f \approx \Omega_\mathrm{m}^{0.55}$ is the dimensionless growth rate in \ac{LCDM}~\citep{Bouchet_1995,Wang_1998} and $b_1$ is the linear galaxy bias. We further include $\Vext$, with uniform prior in both magnitude and direction. Since the reconstruction captures the large-scale velocity field, we assume that residual small-scale motions are described by a diagonal covariance matrix with constant variance $\sigma_v^2 + \sigma_{cz}^2$ as previously. Note that in this case, the explicit modelling of the field-level velocity field renders $\mathbf{\Sigma}_{\Lambda\mathrm{CDM}}$ unnecessary; we investigate the effect of a possible residual covariance below the resolution of~\citetalias{Carrick_2015} in~\cref{sec:vpec_cov}. The reconstruction of~\citetalias{Carrick_2015} was also employed by~\citetalias{Kenworthy_2022}. Further details of the reconstruction are provided in \cref{sec:Carrick_reconstruction}.
    \item Our final, fiducial and most sophisticated (and most realistic) model is based on~\Manticore~\citep{McAlpine_2025}, a density and velocity field reconstruction of the local Universe derived from the \ac{BORG} algorithm~\citep{Jasche_2013,Lavaux_2016,Jasche_2019,Stopyra_2023}. It is shown in~\citet{VF_olympics,McAlpine_2025} that this is the most accurate among all velocity fields currently in the scientific literature. Unlike~\citetalias{Carrick_2015}, \Manticore\ provides posterior samples of the initial conditions, thereby quantifying the reconstruction uncertainty of the local large-scale structure. We use 30 independent posterior samples, each resimulated at higher resolution with an $N$-body code. We describe \Manticore\ further in~\cref{sec:Manticore_reconstruction}. Importantly, we work in real-space rather than redshift-space, thereby avoiding the triple-valued regions that arise when mapping from redshift to real-space distance. While the \ac{BORG} posterior samples have a resolution of $2.6~\Mpch$ in the initial conditions, the \Manticore\ simulations are run at a higher resolution with small-scale modes added below the inference grid scale, over which we then marginalise, effectively accounting for the likely range of smaller-scale density field realisations.
    The \Manticore\ present-day fields extrapolate below the initial condition grid scale to accurately pinpoint positions of galaxy clusters (see Figure~9 of~\citealt{McAlpine_2025}).
    Optionally, we also allow $\sigma_v$ to vary with the local density, following~\cref{eq:sigma_v_density}, to capture any residual small-scale velocity dispersion associated with galaxy clusters. However, this makes little difference and our fiducial model is that of constant $\sigma_v$. As in~\citetalias{Carrick_2015}, we evaluate the \ac{LOS} density and radial velocity, introduce an additional $\Vext$ velocity vector, and model all residual velocities with an uncorrelated, constant $\sigma_v$ term. It is this final model that provides our headline posterior for $H_0$.
\end{enumerate}

\begin{table*}
    \centering
    \begin{tabular}{lp{7cm}p{6cm}}
    \textit{Parameter} & \textit{Description} & \textit{Prior} \\
    \toprule
    $H_0$ & Hubble constant & $\pi(H_0) = \mathcal{U}(10, 100)\kmsecMpc$ \\
    \addlinespace
    \multicolumn{3}{l}{\textbf{Cepheid period--luminosity relation}} \\
    \midrule
    $M_W$ & Absolute magnitude zero-point & $\pi(M_W) = \mathcal{U}(-7,\,-5)$\\
    $b_W$ & period--luminosity slope & $\pi(b_W) = \mathcal{U}(-6,\,0)$ \\
    $Z_W$ & Metallicity correction & $\pi(Z_W) = \mathcal{U}(-2,\,2)$ \\
    $\Delta_\mathrm{ZP}$ & Zero-point offset between ground- and space-based LMC photometry & $\pi(\Delta_\mathrm{ZP}) = \mathcal{N}(0,\,0.1)$ \\
    \addlinespace
    \multicolumn{3}{l}{\textbf{Physical distances}} \\
    \midrule
    $\mu_i$ & Distance moduli of 35 Cepheid host galaxies, LMC, M31 and NGC\,4258 & $\pi(\mu_i) = \pi(r_i) \left|\dv{r}{\mu}\right|_{\mu_i}$ with $\pi(r_i) \propto r_i^2$\\
    $\alpha_{\rm low}$, $\alpha_{\rm high}$, $\ln \rho_t$, $\Delta_{\ln\rho}$ & \Manticore\ galaxy bias parameters & $\pi(\alpha_{\rm low}) = \mathcal{N}(1, 1)$, $\pi(\alpha_{\rm high}) = \mathcal{N}(0.5, 1)$, $\pi(\ln\rho_t) = \mathcal{N}(0, 2)$ and $\pi(\Delta_{\ln\rho}) = \mathcal{N}(0.5,0.5)$ truncated to $0.05 \le \Delta_{\ln\rho} \le 3$\\
    \addlinespace
    \multicolumn{3}{l}{\textbf{Peculiar velocity modelling}} \\
    \midrule
    $\sigma_v$ & Small-scale velocity dispersion & $\pi(\sigma_v) \propto 1/\sigma_v$ \\
    $\sigma_{v,\mathrm{low}},\,\sigma_{v,\mathrm{high}},\,\ln \rho_{\sigma_v},\,k_{\sigma_v}$ & Density-dependent dispersion parameters of~\cref{eq:sigma_v_density} (optional) & $\pi(\sigma_{v,\mathrm{low}}) = \pi(\sigma_{v,\mathrm{high}}) = \text{Maxwell}(\text{scale}=200\kmsec)$, $\pi(\ln\rho_{\sigma_v}) = \mathcal{N}(1, 5)$, $\pi(k_{\sigma_v}) = \mathcal{N}(1, 1)$ truncated at $k_{\sigma_v} \ge 0$ \\
    $\Vext$ & Constant (external) flow vector & $\pi(|\Vext|) = \mathcal{U}(0, 1000)\kmsec$  and uniform in sky direction\\
    $A$ & Scaling of the $\Lambda$CDM velocity covariance matrix & $\pi(A) = \mathcal{N}(1,0.5)$ truncated at $A \geq 0$ \\
    $\beta$ & \protect\cite{Carrick_2015} velocity field scaling parameter & $\pi(\beta) = \mathcal{N}(0.43,0.02)$ \\
    \addlinespace
    \multicolumn{3}{l}{\textbf{Selection function modelling}} \\
    \midrule
    $M_B$ & Standardised \ac{SN} absolute magnitude & $\pi(M_B) = \mathcal{U}(-22, -18)$ \\
    \bottomrule
    \end{tabular}
    \caption{Free parameters of the Bayesian hierarchical model described in~\cref{sec:methodology}, with their corresponding priors. The inclusion of specific peculiar velocity parameters depends on the chosen velocity modelling approach; see~\cref{sec:PV_models}. $\mathcal{U}(a, b)$ denotes a uniform prior over the interval $[a, b]$, while $\mathcal{N}(\mu, \sigma)$ denotes a normal prior with mean $\mu$ and standard deviation $\sigma$. The width of the $\Delta_{\rm ZP}$ prior follows~\protect\citet{Riess_2022}.}
    \label{tab:free_parameters}
\end{table*}


\subsection{Selection function modelling}\label{sec:method_selection_function}

\subsubsection{General approach}

An important systematic uncertainty is the manner in which the sample was selected, in particular the impact of ``unobserved'' data (see e.g.~\citealt{gelman04,Kelly_2007,Kelly_2008,Messenger_2013}). Here, we begin more generally with the problem of inference from a flux- or redshift-limited survey, largely following~\citet{Kelly_2008}. To make things more concrete, this formalism will be used notably to describe the population of Cepheid hosts. It is nonetheless fairly general and we will keep it that way till Section~\ref{sec:SN_magnitude}.

Let ${\bm{d}}_i$ denote some observed data vector for the $i$\textsuperscript{th} source, with source parameters $\bm{\theta}_i$ drawn from a population characterised by $\bm{\Lambda}$ (e.g.~the luminosity function or the standardised \ac{SN} absolute magnitude).
The vector $\bm{\theta}_i$ contains the true sky position and $\bm{d}_i$ the observed one.
We typically take the angular observation as noiseless, yielding a Dirac-delta likelihood.
The sky-position prior, more naturally cast jointly with the distance prior as a prior on the 3D position, nevertheless enters the generative process.
We define $S_i$ as the detection indicator, where $S_i = 1$ if the $i$\textsuperscript{th} source is detected and $S_i = 0$ otherwise. If those observations are independent, apart through the population parameter, the likelihood of all data $\bm{d}$ is a product over the likelihoods of the $n$ detected and $N-n$ undetected objects multiplied by their detection indicators. We must multiply by a binomial coefficient $C_n^N \equiv N!/(n!(N-n)!)$ to account for the number of ways of selecting $N-n$ undetected objects out of $N$:
\begin{equation}\label{eq:sel_full}
\begin{split}
    \mathcal{L}(\bm{d} \mid \bm{\Lambda},\,N)
    =
    C_n^N
        \prod_{i\in\mathcal{A}_{\rm obs}} p(S_i=1 \mid \bm{d}_i) \mathcal{L}(\bm{d}_i \mid \bm{\Lambda}) \times \\
        \prod_{j\in\mathcal{A}_{\rm mis}} p(S_j=0 \mid \bm{d}_j) \mathcal{L}(\bm{d}_j \mid \bm{\Lambda}),
\end{split}
\end{equation}
where $\mathcal{A}_{\rm obs}$ and $\mathcal{A}_{\rm mis}$ are the set of observed and unobserved sources respectively. $N$, the total number of galaxies that would have been detected were there no selection, acts as a nuisance parameter in this likelihood. We can now integrate Eq.~\eqref{eq:sel_full} over the unobserved data to produce the observed data likelihood:
\begin{equation}
\begin{aligned}
    \mathcal{L}(\bm{d}_{\rm obs} \mid \bm{\Lambda},\,N) = C_n^N\, &\prod_{i\in\mathcal{A}_{\rm obs}} p(S_i=1 \mid \bm{d}_i) \mathcal{L}(\bm{d}_i \mid \bm{\Lambda}) \\
    \times &\prod_{j\in\mathcal{A}_{\rm mis}} \int \dd \bm{d}_j\,p(S_j=0 \mid \bm{d}_j) \mathcal{L}(\bm{d}_j \mid\bm{\Lambda}) \\
    \propto C_n^N [p(S=0 &\mid \bm{\Lambda})]^{N-n} \prod_{i\in\mathcal{A}_{\rm obs}} p(S_i=1 \mid \bm{d}_i) \mathcal{L}(\bm{d}_i \mid \bm{\Lambda}),
\end{aligned}
\end{equation}
where we have used the fact that
\begin{equation}
    p(S=0 \mid \bm{\Lambda}) \equiv
    \int\dd \bm{d}\;p(S=0\mid \bm{d}) \mathcal{L}(\bm{d}\mid\bm{\Lambda}),
\end{equation}
because the integration over $\bm{d}_j$ makes the $j$ indices irrelevant.
The joint posterior of $\bm{\Lambda}$ and $N$ is
\begin{equation}
\begin{split}
    \mathcal{P}(\bm{\Lambda},\,N \mid \bm{d}_{\rm obs})
    &\propto
    \pi(\bm{\Lambda}) \pi(N) C_n^N [p(S=0\mid\bm{\Lambda})]^{N-n}\\
    &\times \prod_{i\in\mathcal{A}_{\rm obs}} p(S_i=1 \mid \bm{d}_i) \mathcal{L}(\bm{d}_i\mid\bm{\Lambda}),
\end{split}
\end{equation}
where $\pi(\bm{\Lambda})$ and $\pi(N)$ are the priors on $\bm{\Lambda}$ and $N$, respectively. At this point it is useful to marginalise over $N$ to obtain the marginal posterior $\mathcal{P}(\bm{\Lambda} \mid \bm{d}_{\rm obs})$. For scale-invariance and computational simplicity, we choose a log-uniform prior on $N$, $\pi(N) \propto 1/N$. We then obtain the marginal by summing over all possible values of $N$:
\begin{multline}
    \mathcal{P}(\bm{\Lambda}\mid \bm{d}_{\rm obs})
    \propto\\
    \pi(\bm{\Lambda}) \left[\prod_{i\in\mathcal{A}_{\rm obs}} p(S_i=1 \mid \bm{d}_i) \mathcal{L}(d_i \mid \bm{\Lambda})\right] \sum_{N=n}^\infty \frac{C_n^N [p(S=0\mid\bm{\Lambda})]^{N-n}}{N}\\
    \propto \pi(\bm{\Lambda}) \left[p(S=1\mid\bm{\Lambda})\right]^{-n} \left[\prod_{i\in\mathcal{A}_{\rm obs}} p(S_i=1 \mid \bm{d}_i) \mathcal{L}(\bm{d}_i\mid\bm{\Lambda})\right] \\
    \quad\times\sum_{N=n}^\infty C_{n-1}^{N-1} \left[p(S=0\mid \bm{\Lambda})\right]^{N-n}\left[p(S=1\mid\bm{\Lambda})\right]^{n},
\end{multline}
where the second proportionality follows from multiplying and dividing by $p(S=1\mid\bm{\Lambda})^n$ and utilising $C_n^N = C_{n-1}^{N-1} (N/n)$. The probability $p(S=1\mid\bm{\Lambda})$ follows from its definition: $p(S=1\mid\bm{\Lambda}) = 1 - p(S=0\mid\bm{\Lambda})$. We derive an expression herein below. The sum in the second proportionality is exactly the expression for the negative binomial distribution as a function of $N$, which must equal unity by conservation of probability when all possible values are summed over. This produces the final result
\begin{equation}\label{eq:posterior_with_selection}
    \begin{split}
    \mathcal{P}(\bm{\Lambda}\mid \bm{d}_{\rm obs}) &\propto \pi(\bm{\Lambda}) [p(S=1\mid\bm{\Lambda})]^{-n} \\
    &\quad\times \prod_{i\in\mathcal{A}_{\rm obs}} p(S_i=1 \mid \bm{d}_i) \mathcal{L}(\bm{d}_i\mid\bm{\Lambda}).
    \end{split}
\end{equation}
We note that the second term on the right hand side must be constant for all data elements. If the posterior on $N$ is desired, it can be determined from the other marginal
\begin{equation}
    p(N\mid n,\,\bm{\Lambda}) = C_{n-1}^{N-1} [p(S=1\mid\bm{\Lambda})]^n [p(S=0\mid\bm{\Lambda})]^{N-n},
\end{equation}
which can be further marginalized over $\bm{\Lambda}$ as
\begin{equation}
    p(N \mid n) = \int \dd \bm{\Lambda}\,p(N\mid n,\,\bm{\Lambda}) \pi(\bm{\Lambda}),
\end{equation}
or equivalently estimated directly from the posterior samples of $\bm{\Lambda}$.

We use the marginalised posterior of~\cref{eq:posterior_with_selection}, where the selection is entirely encapsulated through the $p(S = 1 \mid \bm{\Lambda})$ term, which represents the fraction of detected samples from the total population.
It is calculated as
\begin{multline}\label{eq:prob_detection}
    p(S = 1 \mid \bm{\Lambda})=\\
        \iint \dd \dpred\;\dd \bm{\theta}\,p(S = 1 \mid \dpred)
        \mathcal{L}(\dpred \mid \bm{\theta},\,\bm{\Lambda}) \pi(\bm{\theta} \mid \bm{\Lambda}),
\end{multline}
where $\dpred$ denotes the predicted vector data (e.g.~the redshift or apparent magnitude) and $\bm{\theta}$ the unknown parameters of the source (e.g.~the 3D position).
The first term in the integrand is the detection indicator, expressed as a function of $\dpred$.
The second term is the likelihood of the $\dpred$, and the third term is the prior on the source parameters. We note that this probability is independent of individual sources through the integration over $\dpred$ and $\bm{\theta}$.

For a catalogue with a hard detection limit $\dlim$ (e.g.~a flux or redshift limit), the detection indicator is given by
\begin{equation}\label{eq:heaviside_selection}
    p(S = 1 \mid d_{\rm pred},\,\bm{\Lambda}) =
    \begin{cases}
    1\quad\mathrm{if}& d_{\rm pred} < \dlim, \\
    0\quad\mathrm{if}& d_{\rm pred} \geq \dlim,
    \end{cases}
\end{equation}
Furthermore, if we have a Gaussian likelihood for the data with uncertainty $\sigma_d$, and no angular selection, then the probability of detection in~\cref{eq:prob_detection} simplifies to
\begin{equation}\label{eq:prob_detection_with_threshold}
    p(S = 1 \mid \bm{\Lambda})
    =
    \int \dd \bm{\theta}\,\Phi\left(\frac{\dlim - d_{\rm pred}}{\sigma_d}\right) \pi(\bm{\theta} \mid \bm{\Lambda}),
\end{equation}
where $d_{\rm pred}$ is a function of $\bm{\theta}$ and $\bm{\Lambda}$, and $\Phi(x)$ is the \ac{CDF} of the standard normal distribution, defined as
\begin{equation}\label{eq:CDF_standard_normal}
    \Phi(x) = \frac{1}{\sqrt{2\pi}} \int_{-\infty}^x e^{-t^2/2} \dd t.
\end{equation}
In practice, we model a smooth probability of selection as
\begin{equation}
    p(S = 1 \mid d_{\rm obs},\, \bm{\Lambda})
    = \Phi\!\left(\frac{d_{\rm lim} - d_{\rm obs}}{\sigma_{\rm sel}}\right),
\end{equation}
where $d_{\rm lim}$ is the truncation point and $\sigma_{\rm sel}$ sets the smoothness of the transition. For $d_{\rm pred} \ll d_{\rm lim}$ the probability tends to unity, while for $d_{\rm pred} \gg d_{\rm lim}$ it falls to zero. If $\sigma_{\rm sel} \rightarrow 0$, then this would reduce to~\cref{eq:heaviside_selection}. Within~\cref{eq:prob_detection}, the integral of a Gaussian \ac{CDF} with a Gaussian density (assuming a Gaussian likelihood) evaluates to another \ac{CDF}:
\begin{equation}
\begin{split}
    \int \mathrm{d}d_{\rm obs}\,
    &\Phi\!\left(\frac{d_{\rm lim} - d_{\rm obs}}{\sigma_{\rm sel}}\right)
    \mathcal{N}(d_{\rm obs};\, d_{\rm pred},\, \sigma_d)\\
    &=
    \Phi\!\left(\frac{d_{\rm lim} - d_{\rm pred}}
    {\sqrt{\sigma_{\rm sel}^2 + \sigma_d^2}}\right).
\end{split}
\end{equation}
As expected, if $\sigma_{\rm sel}$ is much smaller than $\sigma_d$, then the smooth selection approaches the expression in~\cref{eq:prob_detection_with_threshold}. Below, we consider the regime where they are comparable.

We now calculate the selection probability separately for \ac{SN} magnitude-selected and host-galaxy redshift-selected samples. We verify on mock data that the inference of $H_0$ is unbiased when modelling selection in this way, as detailed in~\cref{sec:mock_data_bias_tests}.

\subsubsection{Supernova magnitude selection}\label{sec:SN_magnitude}

For \ac{SN} apparent magnitude selection we have that
\begin{equation}\label{eq:prob_detection_with_threshold_flux}
    p(S = 1 \mid M_B)
    =
    \int \dd^3\bm{r}\,\pi(\bm{r})\,\Phi\left(\frac{m_{\rm SN}^{\lim} - m_{\rm SN}^{\rm pred}}{\sqrt{\tilde{\sigma}_{\rm SN}^2 + \sigma_{\rm SN}^2}}\right),
\end{equation}
where $m_{\rm SN}^{\lim}$ is the \ac{SN} apparent magnitude threshold,
\begin{equation}
    m_{\rm SN}^{\rm pred} = \mu(r) + M_B
\end{equation}
is the predicted apparent magnitude of a \ac{SN} at distance $r$ with absolute magnitude $M_B$, $\tilde{\sigma}_{\rm SN}$ is the selection smoothness term, and $\sigma_{\rm SN}$ is the uncertainty in \ac{SN} apparent magnitude. We set $\tilde{\sigma}_{\rm SN}$ to be 0.15~mag and $\sigma_{\rm SN}$ to the average square root of the diagonal of the \ac{SN} covariance (approximately $0.13~\mathrm{mag}$), though the inference is not very sensitive to either choice.
In \cref{eq:prob_detection_with_threshold_flux}, the integral over the observed sky position has been collapsed by the delta-function angular likelihood, which fixes the observed sky position to the true one.
Because the selection probability marginalises over all possible sources, we must integrate over the full volume.

Assuming no inhomogeneous Malmquist bias ($n_g \propto \mathrm{const.}$), the radial distance and sky position are independent and the angular prior is constant over the sphere.
Writing the spherical volume element as $\dd^3\bm{r} = r^2 \dd r \dd\bm{\Omega}$, the angular integral yields unity and the radial integration retains the Jacobian $r^2$, so the integral reduces to a 1D integral over $r$ with measure $r^2 \dd r$.
At small physical distances, the distance modulus is related to distance (expressed in Mpc) as $\mu \approx 5 \log (r/\text{1 Mpc}) + 25$, and the detection probability simplifies to
\begin{equation}\label{eq:prob_detection_with_threshold_flux_simple}
    p(S = 1 \mid M_B) \propto 10^{-3 M_B / 5}.
\end{equation}
This result is independent of $\sigma_{\rm SN}$. In the expression above we wrote the dependence as $10^{-3M_B/5}$. More precisely, the relevant factor is $10^{3(M_B - m_{\rm SN}^{\lim})/5}$, which depends on the difference between $M_B$ and the limiting magnitude $m_{\rm SN}^{\lim}$. Since the posterior dependence on $m_{\rm SN}^{\lim}$ is separable, it can be marginalised out and absorbed into the overall proportionality constant together with $\sigma_{\rm SN}$. We can also consider the case when the selection is a smooth function of the magnitude. Going back to~\cref{eq:prob_detection}, we have that
\begin{multline}
    p(S = 1 \mid M_B)\propto\\
    \int \dd r \: r^2\,\dd m_{\rm SN}^{\rm pred}\,
    p(S = 1 \mid m_{\rm SN}^{\rm pred})
    \mathcal{L}(m_{\rm SN}^{\rm pred} \mid M_B,\,r).
\end{multline}
Through a similar change of variables we can pull out the dependence on $M_B$ outside of the integral, so that $p(S = 1 \mid M_B) \propto 10^{-3 M_B / 5}$ and the shape of the selection function only changes the normalisation constant such that the posterior is independent of it.

For runs assuming an underlying local Universe reconstruction, $\pi(\bm{r})$ is proportional to the source number density, which depends on the underlying density field, and we evaluate the selection integral as
\begin{equation}\label{eq:selection_normalisation_SN}
    p(S = 1 \mid M_B) \propto \int \dd^3\bm{r}\,n_g(\bm{r})\,\Phi\left(\frac{m_{\rm SN}^{\lim}-\mu(\bm{r};H_0)-M_B}{\sqrt{\tilde{\sigma}_{\rm SN}^2+\sigma_{\rm SN}^2}}\right),
\end{equation}
evaluated as a sum over all voxels inside the reconstruction volume.
The integration volume must extend far enough that voxels beyond it receive vanishing weight from the \ac{CDF}.
Using $n_g(\bm{r})$ in place of the normalised $\pi(\bm{r})$ leaves the right-hand side unnormalised, but the same unnormalised measure enters the detected-host distance integrals, so the missing normalisation factor cancels exactly in the posterior.

When accounting for the selection in \ac{SN} apparent magnitude, the inference is no longer independent of the \acp{SN} and depends explicitly on $M_B$. The intrinsically brighter the \acp{SN}, the more of them pass the selection threshold and are observed at greater distances. This behaviour holds regardless of \acp{SN} being observed at much greater distances than those considered here; only the host galaxies included in the sample of 35 matter. Moreover, their absolute magnitudes must be modelled self-consistently with the inferred distances to the Cepheid host galaxies and with the \ac{SN} apparent magnitudes used in the selection. We treat $M_B$ as a free parameter and sample it explicitly, and therefore we also forward-model the apparent magnitudes of \acp{SN} in Cepheid host galaxies with the likelihood term to constrain it,
\begin{equation}\label{eq:SN_magnitude_likelihood}
    \mathcal{L}(\bm{m}_{\rm SN} \mid \bm{m}_{\rm SN}^{\rm pred}) = \mathcal{N}(\bm{m}_{\rm SN}; \bm{m}_{\rm SN}^{\rm pred}, \mathbf{\Sigma}_{\rm SN}),
\end{equation}
where $\bm{m}_{\rm SN}$ are the bias-corrected \ac{SN} apparent magnitudes from the SH0ES sample, $\bm{m}_{\rm SN}^{\rm pred}$ are the predicted \ac{SN} apparent magnitudes, and $\mathbf{\Sigma}_{\rm SN}$ is the covariance matrix of the \ac{SN} apparent magnitudes. Some of the 35 Cepheid host galaxies contain multiple \acp{SN}; however, in such cases we select only the brightest \ac{SN} per host. In deriving~\cref{eq:prob_detection_with_threshold_flux}, we assumed the \ac{SN} observations are independent, which is not valid due to correlations introduced by the standardisation procedure (i.e.~non-zero off-diagonal elements in the covariance matrix). In contrast, the corresponding likelihood in~\cref{eq:SN_magnitude_likelihood} uses the covariance matrix. We verify that neither the downsampling of \acp{SN} to one per galaxy nor setting all the off-diagonal elements of the \ac{SN} magnitude covariance matrix to zero affects the $H_0$ posterior distribution.

\subsubsection{Redshift-limited selection}\label{sec:redshift_selection}

For a redshift-limited survey,
\begin{equation}\label{eq:prob_detection_with_threshold_redshift}
    p(S = 1 \mid H_0)
    =
    \int \dd^3\bm{r}\,\pi(\bm{r})\,\Phi\left(\frac{c \zlim - c z^{\rm pred}}{\sqrt{\tilde{\sigma}_v^2 + \sigma^2_v}}\right),
\end{equation}
where $z^{\rm pred}$ is the predicted redshift of a galaxy at 3D position $\bm{r}$, and $\tilde{\sigma}_v$ is the redshift truncation width, assumed to be $300\kmsec$, though the inference is not sensitive to this choice. For illustration, if peculiar velocities and inhomogeneous Malmquist bias are neglected and $\sigma_v$ is fixed, then at low redshift, where $c\zCMB \approx H_0 r$, the detection probability for a redshift-selected survey simplifies to
\begin{equation}\label{eq:prob_detection_with_threshold_redshift_simple}
    p(S = 1 \mid H_0) \propto H_0^{-3},
\end{equation}
per host (and these probabilities multiply, yielding an effective $H_0^{+105}$ dependence, which can shift the peak significantly), with the redshift cut-off absorbed into the proportionality constant.
For runs assuming an underlying local Universe reconstruction, we follow a treatment analogous to \cref{eq:selection_normalisation_SN}.

As with the \ac{SN} magnitude selection, the detection probability in~\cref{eq:prob_detection_with_threshold_redshift} assumes mutually independent source redshifts, corresponding to a diagonal covariance matrix with entries of $\sigma_v^2$. This assumption holds for \Manticore\ where we take each realisation (without covariance) independently and combine them self-consistently, and also for~\citetalias{Carrick_2015} which provides a single realisation. We assume any residual correlations to be negligible. However, it fails when adopting the \ac{LCDM} peculiar velocity covariance matrix. As demonstrated in~\cref{fig:host_covariance}, the covariance matrix for the 35 host galaxies exhibits significant off-diagonal correlations.

To ``correct'' for the loss of information due to correlated samples, we estimate the effective number of samples: given a generic covariance matrix $\mathbf{C}$ of dimension $N\times N$ we can calculate the effective rank as the exponential of the Shannon entropy of its normalised eigenvalues,
\begin{equation}\label{eq:matrix_effective_rank}
    \mathrm{N_{eff}}(\mathbf{C}) = \exp\left[ -\sum_{i=1}^{N} p_i \log p_i \right],
\end{equation}
where $p_i = \lambda_i / \sum_{j=1}^N \lambda_j$, and $\lambda_i$ are the eigenvalues of $\mathbf{C}$~\citep{Roy_2007}. $\mathrm{N_{eff}}(\mathbf{C})$ provides a continuous measure of the matrix's effective dimensionality, with $\mathrm{N_{eff}}(\mathbf{C}) = N$ for a diagonal covariance matrix. This produces $N_{\mathrm{eff}} = 21.0$ for the \ac{LCDM} peculiar velocity covariance matrix for the 35 host galaxies. Accordingly, we down-weight the contribution of~\cref{eq:prob_detection_with_threshold_redshift} in the model likelihood by a factor of $N_{\mathrm{eff}} / N \approx 0.57$ for that peculiar velocity model. However, we stress that this correction is only approximate. No such correction is required for the \ac{SN} magnitude selection used here; the resulting constraints on $H_0$ are unchanged when the corresponding magnitude covariance is diagonalised in the likelihood.

\subsubsection{Joint supernova magnitude- and redshift-limited selection}\label{sec:SN_mag_redshift_selection}

Although not a fiducial model, we also consider the case of selection on \emph{both} \ac{SN} host magnitude and galaxy redshift. In this case, the selection probability is given by
\begin{equation}\label{eq:prob_detection_with_joint_threshold}
\begin{split}
    p(&S = 1 \mid H_0, M_B) =\\
    &\int \dd^3\bm{r}\,\pi(\bm{r})\,
    \Phi\left(\frac{m_{\rm SN}^{\lim} - m_{\rm SN}^{\rm pred}}{\sqrt{\tilde{\sigma}_{\rm SN}^2 + \sigma_{\rm SN}^2}}\right)
    \Phi\left(\frac{c \zlim - c z^{\rm pred}}{\sqrt{\tilde{\sigma}_v^2 + \sigma^2_v}}\right),
\end{split}
\end{equation}
which is computed analogously to the previous cases, except that we have a product of the \acp{CDF}. For this selection, when modelling the peculiar velocities with the \ac{LCDM} covariance (but crucially not with either \Manticore\ or \citetalias{Carrick_2015}), we use the same correction for the number of effective samples as in~\cref{sec:redshift_selection}.


\section{Results}\label{sec:results}

As a preliminary to our $H_0$ inference, in~\cref{sec:Cepheid_only_distances} we infer the host galaxy distances without using their redshifts. This enables us to quantify the effect of priors and selection modelling on the inferred distances. Then, in~\cref{sec:Cepheid_only_H0} we add the redshift information to constrain $H_0$ with the full model (i.e. including the part in the dashed box of~\cref{fig:CH0_DAG}).
In~\cref{tab:cepheid_hosts} we provide the inferred Cepheid host distances and the peculiar velocities from both \citetalias{Carrick_2015} and \Manticore.

\subsection{Host distances}\label{sec:Cepheid_only_distances}

\begin{figure}
    \centering
    \includegraphics[width=\columnwidth]{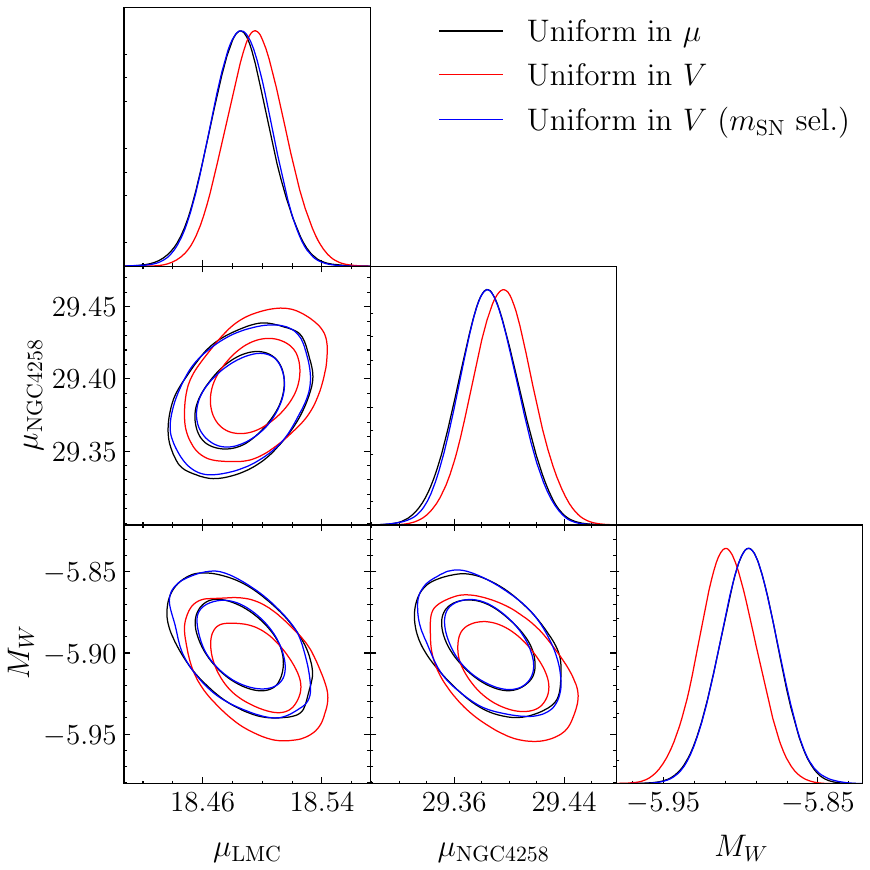}
    \caption{Corner plot of the inferred distance moduli to the \ac{LMC}, NGC\,4258, and the \ac{CPLR} zero-point calibration, from analyses that do not incorporate redshift information. We compare three scenarios: a uniform-in-distance-modulus prior without selection modelling, a uniform-in-volume prior without selection modelling, and a uniform-in-volume prior with \ac{SN} magnitude selection modelled. The uniform-in-volume posterior with \ac{SN} magnitude selection is in close agreement with the uniform-in-distance-modulus calibration. Contours denote $1\sigma$ and $2\sigma$ confidence intervals.}
    \label{fig:anchor_distances}
\end{figure}

We now consider only the Cepheid host distances, calibrated using the \ac{CPLR} and two geometric anchors (\ac{LMC} and NGC\,4258). We examine four modelling scenarios:
\begin{enumerate}
	    \item a uniform prior in distance modulus without selection modelling (a baseline reference);
	    \item a uniform prior in volume without selection modelling;
	    \item a uniform prior in volume with modelling of \ac{SN} magnitude selection;
	    \item a uniform prior in volume with modelling of host-redshift selection.
\end{enumerate}
The uniform-in-volume prior, reflecting the fact that the Universe is three-dimensional, yields unbiased distance-ladder inference. In contrast, the uniform-in-distance-modulus prior has no physical motivation and is implicitly adopted in~\citetalias{Kenworthy_2022} and the SH0ES analysis~\citep{Riess_2022} where maximum-likelihood parameters are estimated by minimising a joint $\chi^2$ statistic across the distance ladder including the host galaxy distance moduli as free parameters. The uniform-in-volume prior follows $p(r_i) \propto r_i^2$, while the uniform-in-distance-modulus prior corresponds to $p(r_i) \propto 1 / r_i$. Since the uniform-in-distance-modulus prior biases the inferred distances low relative to a uniform-in-volume prior, for fixed redshifts it biases $H_0$ high. This is discussed in more detail and generality in~\citet{Desmond_2025}. \cref{fig:anchor_distances} shows the inferred distance moduli to the \ac{LMC} and NGC\,4258 along with the Cepheid zero-point $M_W$. We compare our results to the SH0ES analysis~\citep{Riess_2022}. The no-selection, uniform-in-distance-modulus prior (red contours) produces results in excellent agreement with the SH0ES analysis, as expected. In contrast, the no-selection, uniform-in-volume prior (blue contours) yields larger inferred distances and a brighter Cepheid absolute magnitude zero-point. Incorporating \ac{SN} magnitude selection with a uniform-in-volume prior shifts the inferred distances to the \ac{LMC} and NGC\,4258 and Cepheid zero-point back towards those obtained under a uniform-in-distance-modulus prior without selection modelling (see~\cref{fig:anchor_distances}).

We verify that in any case, the inferred Cepheid-zero-point $M_W$ is not in any tension with the \ac{MW} calibration, which was used as a constraint in the model. Similarly, the inferred distances to the \ac{LMC} and NGC\,4258 remain consistent with the imposed external calibration. Moreover,~\cref{fig:anchor_distances} also shows that, in the absence of selection modelling, the choice of prior shifts the distance to NGC\,4258 but leaves the distance to the \ac{LMC} unchanged. This suggests that the \ac{MW} Cepheid zero-point calibration, with precise parallax distances, is unaffected by the prior choice.

\begin{figure}
    \centering
    \includegraphics[width=\columnwidth]{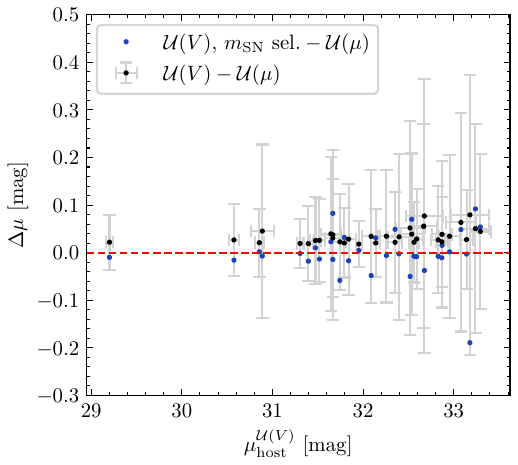}
    \caption{Comparison of distances to the 35 Cepheid host galaxies between a Cepheid-only distance inference using a uniform-in-volume prior without selection (black), or with \ac{SN} magnitude selection (blue), relative to a uniform-in-distance-modulus prior without selection modelling. The first yields an average distance shift of $0.035~\mathrm{mag}$, corresponding to an approximately $1.6$ per cent upward bias in $H_0$ if a uniform-in-distance-modulus prior were used instead, while the second yields no systematic offset. The error bars represent $1\sigma$ uncertainties.}
    \label{fig:mu_host}
\end{figure}

In~\cref{fig:mu_host} we extend the comparison of inferred distances to all 35 Cepheid host galaxies. In the absence of selection modelling, we find that the uniform-in-volume prior yields, on average, a distance modulus larger by $0.035 \pm 0.136~\mathrm{mag}$ compared to the uniform-in-distance-modulus prior. Thus, using a uniform-in-distance-modulus prior instead of uniform-in-volume potentially biases the inferred value of $H_0$ high by approximately $1.6$ per cent. On the other hand, a uniform-in-volume prior with \ac{SN} selection modelled yields distance moduli that are not systematically offset from those obtained with a uniform-in-distance-modulus prior without selection modelling; the mean difference is $0.0014 \pm 0.122~\mathrm{mag}$. From this point onward, we adopt the physically motivated uniform-in-volume prior. We note that in~\cref{sec:Cepheid_only_H0}, when inferring $H_0$ using either \Manticore\ or~\citetalias{Carrick_2015}, we apply this prior to host galaxy distances and account for inhomogeneous Malmquist bias due to source density fluctuations induced by underlying density fluctuations.

\begin{figure*}
    \centering
    \includegraphics[width=\textwidth]{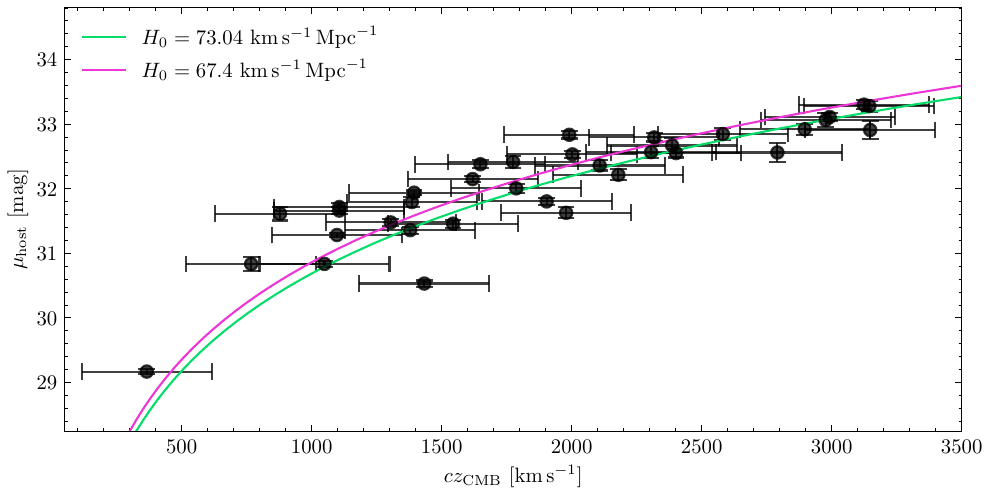}
    \caption{The Hubble diagram for the 35 Cepheid host galaxies, showing the relation between \ac{CMB}-frame redshifts $\zCMB$ and the inferred distance moduli from our analysis assuming a uniform-in-volume prior and a selection in \ac{SN} apparent magnitude. The red curve indicates the predicted distance modulus--redshift relation for the SH0ES best-fit value of $H_0 = 73.0\kmsecMpc$~\protect\citep{Riess_2022}, while the blue curve shows the relation for the \textit{Planck} value $H_0 = 67.4\kmsecMpc$~\protect\citep{Planck_2020_cosmo}. For plotting purposes, we assume a redshift uncertainty corresponding to a velocity dispersion of $250\kmsec$ though the exact value would depend on peculiar velocity modelling. The error bars represent $1\sigma$ uncertainties.}
    \label{fig:mu_host_cz}
\end{figure*}

\subsection{Hubble constant inference}\label{sec:Cepheid_only_H0}

To set the stage for our inference of $H_0$, and visualize the data, we present in~\cref{fig:mu_host_cz} the Hubble diagram showing the relation between the observed redshifts of the 35 host galaxies converted to the \ac{CMB} frame and their inferred distance moduli under the uniform-in-volume prior and the \ac{SN} magnitude selection modelling from the previous subsection. The red and blue lines indicate the predictions for the best-fit SH0ES and \textit{Planck} values of $H_0$, respectively.

We now use the full forward model (see~\cref{fig:CH0_DAG}) to infer $H_0$ as well as the distances by folding in the Cepheid host galaxy redshift information. In~\cref{tab:H0_PV_vs_selection}, we tabulate the inferred values of $H_0$ for various peculiar velocity and selection function models considered in Sec.~\ref{sec:methodology}, all assuming a uniform-in-volume prior on the host galaxy distances. \Cref{fig:H0_comparison} compares the inferred $H_0$ for the various selection models while modelling peculiar velocities with \Manticore. We present a stacked $H_0$ comparison in~\cref{fig:H0_stacked}.

\subsubsection{Supernova magnitude selection}

We now consider in more detail the case where the sample is assumed to be selected by \ac{SN} apparent magnitude solely (described in~\cref{sec:SN_magnitude}). As outlined in~\cref{sec:PV_models}, our simplest models assume uncorrelated peculiar velocities, either setting them to zero on average, or modelling them with a constant velocity vector $\Vext$ that is sampled jointly. The redshift scatter is described by a free parameter $\sigma_v$ that is also sampled. In these cases, we find $H_0 = 68.9 \pm 1.9$ and $69.3 \pm 1.7\kmsecMpc$, respectively. However, these results should be interpreted with caution, as they rely on highly simplified treatments of peculiar velocities. An alternative and more conservative approach adopts the \ac{LCDM} peculiar velocity covariance matrix (described in~\cref{sec:LCDM_covariance}). This effectively marginalises over all plausible peculiar velocity realisations consistent with the \ac{LCDM} power spectrum. It yields $H_0 = 70.3 \pm 3.0$, or $70.1 \pm 3.3\kmsecMpc$ when including a global scaling factor $A$ for the covariance matrix, with it constrained to $1.2 \pm 0.3$.

We now account for peculiar velocities using local Universe reconstructions. First, we use the~\citetalias{Carrick_2015} reconstruction, with $\Vext$ to capture large-scale flows sourced by structure outside the reconstructed volume. Smaller-scale motions are treated as uncorrelated and modelled with a velocity dispersion $\sigma_v$. As shown in~\cref{fig:H0_comparison}, this model yields $H_0 = 71.4\pm1.6\kmsecMpc$. When using~\citetalias{Carrick_2015}, we find $\sigma_v = 217 \pm 32\kmsec$, $|\Vext| = 187 \pm 82\kmsec$ and $\beta = 0.42\pm0.02$ (sampled with a Gaussian prior with mean 0.43 and standard deviation 0.02, following~\citetalias{Carrick_2015}). To account for \ac{SN} magnitude selection, we infer and marginalise over the standardised \ac{SN} absolute magnitude $M_B$.

Second, we use the \Manticore\ reconstruction, finding $H_0 = 71.1 \pm 1.4\kmsecMpc$. In this setup, the velocity field scaling parameter $\beta$ is fixed to unity (since \Manticore\ is a dark matter resimulation assuming a fixed cosmology) and we find that $\sigma_v = 173 \pm 27\kmsec$ and $|\Vext| = 258 \pm 66\kmsec$. In \cref{fig:Manticore_corner}, we show the posterior distribution of all model parameters besides the 35 host galaxy distances. In particular, the posterior on $M_B$ agrees well with the reported value of $M_B = -19.253$ in Fig.~14 of~\cite{Riess_2022}.
Letting $\sigma_v$ follow the density-dependent form of~\cref{eq:sigma_v_density} yields $H_0 = 71.1 \pm 1.2\kmsecMpc$, almost unchanged from the fiducial value.
We also consider the case where $\beta$ is allowed to vary, imposing a Gaussian prior with mean 1 and standard deviation 0.5, loosely corresponding to varying the assumed value of $\sigma_8$. This produces $H_0 = 71.0\pm1.4\kmsecMpc$ with $\beta = 0.90\pm0.20$, which is in good agreement with the $\beta = 1$ inference. Using \Manticore\ yields the smallest uncertainty on $H_0$ due to its most precise (and accurate;~\citealt{VF_olympics,McAlpine_2025}) determination of the peculiar velocity field. The smaller inferred $\sigma_v$ than~\citetalias{Carrick_2015} also indicates that it more successfully captures local flows.

\subsubsection{Host galaxy redshift selection}

We now consider the case in which the sample of host galaxies is assumed to be selected solely based on observed redshift, as described in~\cref{sec:redshift_selection}. Our two simplest models, both using a diagonal covariance matrix with a free parameter $\sigma_v$, assume either zero-mean peculiar velocities or model them with a constant velocity vector $\Vext$. These yield $H_0 = 73.9 \pm 2.8$ and $73.0 \pm 2.3\kmsecMpc$, respectively. In contrast, assuming zero-mean peculiar velocities but including the \ac{LCDM} peculiar velocity covariance matrix, we find $H_0 = 76.2 \pm 3.1$, or $78.3 \pm 4.0\kmsecMpc$ when allowing for a global scaling factor $A$, which is found to be $1.4 \pm 0.3$. However, these results require an approximate correction to the selection function to account for the effective number of independent samples (see~\cref{sec:redshift_selection}), so the results obtained using the \ac{LCDM} covariance should be interpreted with caution. No such correction is necessary when peculiar velocities are explicitly modelled and the covariance is diagonal.

Then, we again consider our two local Universe reconstruction models. Using the~\citetalias{Carrick_2015} reconstruction, we find $H_0 = 73.8\pm1.7\kmsecMpc$, with $\beta = 0.42 \pm 0.02$, which is in good agreement with SH0ES. In contrast, using the \Manticore\ reconstruction yields $H_0 = 72.5 \pm 1.4\kmsecMpc$. \cref{fig:Manticore_corner} presents the posterior distribution assuming \Manticore\ and redshift-limited selection, showing that the inferred $\Vext$ and $\sigma_v$ are consistent with the \ac{SN} selection case.
Applying the density-dependent $\sigma_v$ model of~\cref{eq:sigma_v_density} gives $H_0 = 72.2 \pm 1.3\kmsecMpc$.
If instead of fixing $\beta$ to unity for \Manticore\ we sample it with a Gaussian prior with mean unity and standard deviation 0.5, we obtain consistent values of $H_0 = 72.5 \pm 1.4~\kmsecMpc$ and $\beta = 1.02 \pm 0.19$.

\subsubsection{Joint magnitude and redshift selection}

\begin{figure}
    \centering
    \includegraphics[width=\columnwidth]{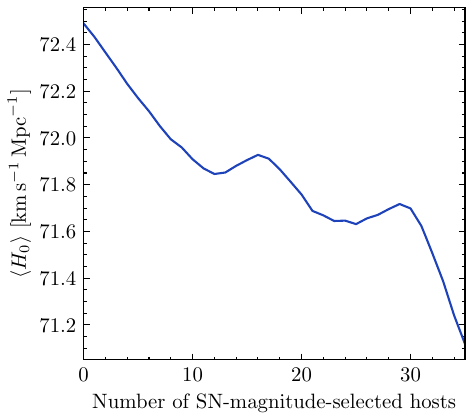}
    \caption{Mean inferred $H_0$ using \Manticore\ for the mixed-selection sweep as the number of host galaxies assigned to the \ac{SN}-magnitude-selected subset is varied from 0 to 35.
    The endpoints correspond to the pure host-redshift-selection and pure \ac{SN}-magnitude-selection models.}
    \label{fig:H0_proportion}
\end{figure}

\begin{figure*}
    \centering
    \includegraphics[width=\textwidth]{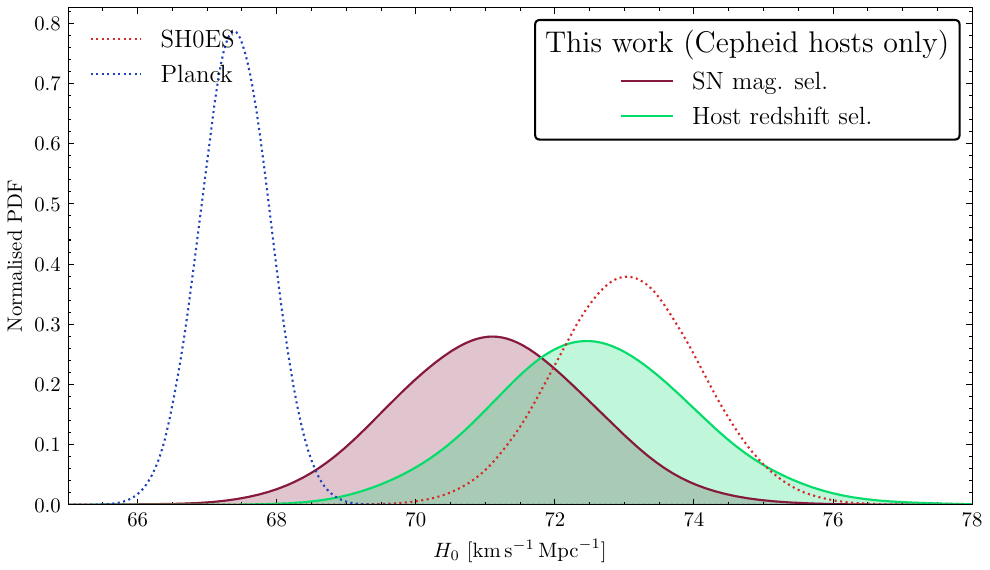}
    \caption{Posterior distributions of $H_0$ for \ac{SN} magnitude selection (maroon) and host galaxy redshift selection (green) when accounting for peculiar velocities using \Manticore~\protect\citep{McAlpine_2025}, compared to \textit{Planck} ($67.4 \pm 0.5\kmsecMpc$;~\protect\citealt{Planck_2020_cosmo}) and SH0ES ($73.0 \pm 1.0\kmsecMpc$;~\protect\citealt{Riess_2022}). The \ac{SN} magnitude- and host-redshift-selection posteriors are mutually consistent and peak below SH0ES. Model variations are listed in~\cref{tab:H0_PV_vs_selection}, and~\cref{fig:H0_stacked} compares our results using both \Manticore\ and~\protect\cite{Carrick_2015} with those from the literature.}
    \label{fig:H0_comparison}
\end{figure*}

\begin{figure*}
    \centering
    \includegraphics[width=\textwidth]{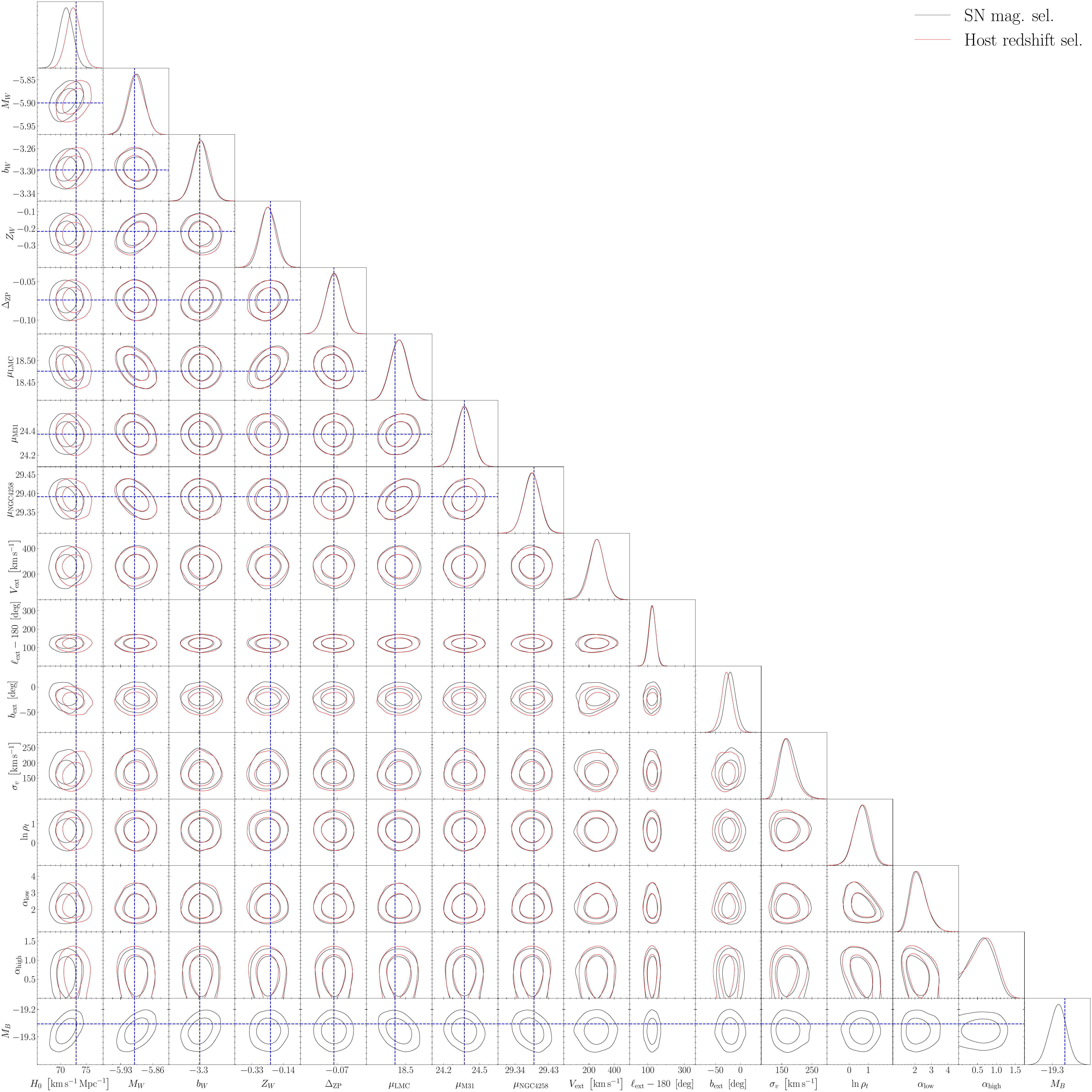}
    \caption{Constraints on all model parameters, except the 35 Cepheid host distances, for our fiducial peculiar velocity model based on the \Manticore\ reconstruction~\protect\citep{McAlpine_2025}, assuming either \ac{SN} magnitude (black) or host redshift selection (red). The blue lines show the best-fit parameter values from the SH0ES analysis~\protect\citep{Riess_2022}. The contours represent $1$ and $2\sigma$ confidence intervals.}
    \label{fig:Manticore_corner}
\end{figure*}

For completeness, we briefly consider mixed selection scenarios in which the sample is interpreted as a concatenation of two subsamples: one selected purely on \ac{SN} magnitude and the other purely on host redshift.
\cref{fig:H0_proportion} shows the mean inferred $H_0$ as a function of the number of galaxies selected by \ac{SN} magnitude, varying this number from zero to 35. The left bound corresponds to purely redshift-based selection and the right bound to purely \ac{SN}-magnitude-based selection. We find an approximately linear scaling between these two limiting cases, with an average $H_0$ uncertainty very close to ${\sim}1.4\kmsecMpc$ in all cases.

\begin{table*}
    \centering
    \begin{tabular}{
        l
        c
        c
    }
    & \multicolumn{2}{c}{\textbf{Selection model}} \\
    \cmidrule(lr){2-3}
    \textbf{Peculiar velocity model}
        & SN mag.
        & Redshift \\
    \midrule
	    No pec. vel., $\sigma_v$
	        & $68.9 \pm 1.9$ & $73.9 \pm 2.8$ \\
	    Constant inferred flow $\Vext$, $\sigma_v$
	        & $69.3 \pm 1.7$ & $73.0 \pm 2.3$ \\
	    $\Lambda$CDM covariance, $\sigma_v^2$
	        & $70.3 \pm 3.0$ & $76.2 \pm 3.1$ \\
	    $\Lambda$CDM covariance, scaling $A$, $\sigma_v^2$
	        & $70.1 \pm 3.3$ & $78.3 \pm 4.0$ \\
	    \citet{Carrick_2015}, $\Vext$, $\sigma_v$
	        & $71.4 \pm 1.6$ & $73.8 \pm 1.7$ \\
	    \textbf{\Manticore, $\Vext$, $\sigma_v$}
	        & $\mathbf{71.1 \pm 1.4}$ & $72.5 \pm 1.4$ \\
	    \textbf{\Manticore, $\Vext$, $\sigma_v(\delta)$}
	        & $71.1 \pm 1.2$ & $72.2 \pm 1.3$ \\
	    \Manticore, $\Vext$, $\sigma_v$, $\beta$
	        & $71.0 \pm 1.4$ & $72.5 \pm 1.4$ \\
    \addlinespace
    \midrule
    \multicolumn{3}{l}{\textit{External comparison}} \\
    \midrule
    CMB~\citep{Planck_2020_cosmo}
        & \multicolumn{2}{c}{$67.4 \pm 0.5$} \\
    SH0ES~\citep{Riess_2022}
        & \multicolumn{2}{c}{$73.0 \pm 1.0$} \\
    SH0ES (with the SMC,~\citealt{Breuval_2024})
        & \multicolumn{2}{c}{$73.2 \pm 0.9$} \\
    SH0ES Cepheid-only (\citealt{Kenworthy_2022})
        & \multicolumn{2}{c}{$72.9 \pm 2.3$} \\
    \bottomrule
    \end{tabular}
	    \caption{Inferred values of the Hubble constant $H_0$ in units of $\mathrm{km}\,\mathrm{s}^{-1}\,\mathrm{Mpc}^{-1}$ for different combinations of peculiar velocity models (rows) and selection models (columns). All results are based on the forward modelling approach described in~\cref{sec:forward_model}. See~\cref{sec:Cepheid_only_H0} for details. We highlight in bold peculiar velocity modelling with \Manticore~\protect\citep{McAlpine_2025}, which we consider our fiducial model. The lower block contains literature constraints for external comparison. All uncertainties are 1$\sigma$ statistical except for that of~\protect\citeauthor{Kenworthy_2022}, which combines statistical and systematic uncertainty.}
    \label{tab:H0_PV_vs_selection}
\end{table*}

\begin{figure}
    \centering
    \includegraphics[width=\columnwidth]{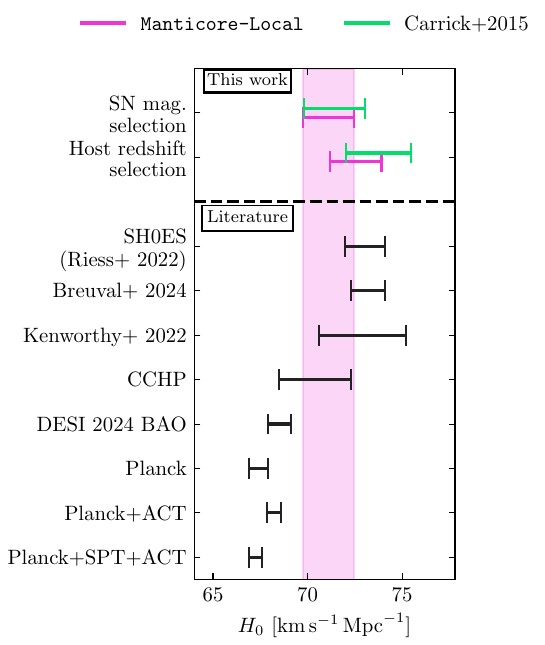}
    \caption{We have shown that a Cepheid-only distance ladder yields $H_0$ values consistent with SH0ES~\protect\citep{Riess_2022}, although using the state-of-the-art \Manticore\ reconstruction to model galaxy bias and account for peculiar velocities gives somewhat lower values unless the host sample is assumed to be redshift-selected.}
    \label{fig:H0_stacked}
\end{figure}


\section{Discussion}\label{sec:discussion}

In this section we discuss the impact of selection function modelling on the inferred $H_0$ (\cref{sec:discussion_selection}) and then compare our results with those of~\citetalias{Kenworthy_2022} (\cref{sec:kenworthy_comparison}).

\subsection{Importance of selection function modelling}\label{sec:discussion_selection}

Typically (or at least ideally), the selection strategy for a survey is well-defined and imposed, for example, on apparent magnitude (i.e.\,a flux limit), reflecting the detector's sensitivity. However, as the sample of Cepheid host galaxies has been assembled over many years as follow-up to nearby \ac{SN} hosts from various proposals, it is unclear which selection is most applicable~\citepalias{Kenworthy_2022}. Nonetheless, given that all galaxies in the sample are relatively nearby, some form of selection must be present. Therefore, we consider two main selection scenarios: a \ac{SN} apparent magnitude limit and a host galaxy redshift limit. \citetalias{Kenworthy_2022} considered these same two cases, arguing for a \ac{SN}-based selection on the grounds that Cepheid hosts are typically targeted for follow-up after a \ac{SN} is observed, with a preference for brighter (nearer) \acp{SN} within which (once selected) Cepheids are always observed.

Given the distributions of \ac{SN} magnitudes and host redshifts shown in~\cref{fig:SH0ES_sample}, we consider selection based solely on \ac{SN} magnitude to be the more plausible scenario. Compared to the \PP\ sample, such a selection can reproduce the Cepheid host redshift distribution, whereas selection on redshift alone fails to reproduce the \ac{SN} magnitude distribution. We briefly explore mixed selection scenarios in~\cref{sec:SN_mag_redshift_selection}, finding that they simply interpolate between the two cases. The effect of Cepheid selection may however also need to be considered. This would require modelling not only the Cepheid population (e.g. period and metallicity distributions) but also Cepheids' spatial distribution within host galaxies, which affects detectability, and propagating the resulting non-trivial covariance matrix. We leave this for future work.

If the sample is selected by \ac{SN} magnitude, neglecting this selection biases the inferred $H_0$ because the selection preferentially retains nearby, brighter sources. Removing this bias requires the explicit treatment of~\cref{sec:SN_magnitude}, which (weakly) couples the inference to the \acp{SN} through the correlation between \ac{SN} and Cepheid apparent magnitudes at fixed host distance. Mock tests in~\cref{sec:mock_data_bias_tests} confirm that this treatment yields unbiased $H_0$. We have also verified that including \ac{SN} apparent magnitudes in the forward model without modelling their selection has no effect on the inferred $H_0$. If the sample is instead assumed selected on Cepheid apparent magnitude or host redshift, the analysis is independent of \ac{SN} data.

\subsection{Comparison to Kenworthy et al.~2022}\label{sec:kenworthy_comparison}

While using the exact same 35 host galaxies, our approach has three principal differences to that of~\citetalias{Kenworthy_2022}:
\begin{enumerate}
    \item We forward model the Cepheid apparent magnitudes, rather than adopt a-priori calibrated distance estimates.
    \item We employ a principled treatment of selection effects, avoiding arbitrary modifications to the prior on object distances (unless considering an effective selection model).
    \item We use a state-of-the-art local Universe reconstruction \Manticore, rather than relying solely on linear-theory reconstructions. This marginalises over the full posterior rather than assuming a Gaussian covariance.
\end{enumerate}
We now compare the methods in detail, showing that neither the redshift- nor distance-limited selection of~\citetalias{Kenworthy_2022} is well justified. A distance-based selection cannot be considered in the first place, since selection must act on the level of an observable quantity, which distance is not.

\subsubsection{Redshift selection}

For redshift selection,~\citetalias{Kenworthy_2022} model the case in which only the host galaxy redshift was effectively used in the selection procedure. It appears that they mean this to imply a prior on the comoving distance $\bm\chi$ of the Cepheid hosts that models homogeneous and inhomogeneous Malmquist bias in the usual way (their Eq.~15), without any further selection modifications. However, they then write ``Tests on simulated data found that the use of this prior tended to bias the recovered distances and $H_0$ high'', and proceed to modify the prior to:
\begin{equation}
    \pi(\bm{\chi}) = \max(\bm{\chi})^\alpha \: \prod_i \frac{\chi_i^2}{\int_0^{\max(\bm{\chi})} \chi^2 \dd \chi},
\end{equation}
where $\bm{\chi}$ is expressed in $h^{-1}\,\mathrm{Mpc}$ and $\alpha$ is set to 60. This entails that:
\begin{enumerate}
    \item The individual $\chi_i^2$ terms are normalised at each step to the current maximum inferred host distance, effectively truncating the $\chi^2$ prior at $\max(\bm{\chi})$.
    \item The prior is not a valid probability density function, as the $\max(\bm{\chi})^\alpha$ factor breaks the normalisation. Moreover, it is not merely an improper prior on distance but it explicitly depends on $h$, which is a separate (and the crucial) parameter.
\end{enumerate}

To quantify the bias on $H_0$ that this produces, we consider the relation between priors on $\chi_i$ and $r_i$, where $r_i$ is physical distance in Mpc such that $\chi_i = r_i h$. If $\pi(\chi_i) = 3 \chi_i^2 / \chi_{\max}^3$, which corresponds to a prior on $\chi_i$ that goes as $\chi_i^2$ but truncated at $\chi_{\max}$, the corresponding prior on $r_i$ is
\begin{equation}
    \pi(r_i) = \pi(\chi_i) \left|\dv{\chi}{r}\right|_{r_i} = \frac{3 r_i^2}{r_{\max}^3},
\end{equation}
where $r_{\max} = \chi_{\max} / h$. Therefore, provided that this term is normalised as it is in~\citetalias{Kenworthy_2022}'s Eq.~17, it does not introduce spurious factors of $h$. However, the factor of $\max(\bm{\chi})^\alpha$ introduces a problem by modifying the ``prior'' on individual distances to
\begin{equation}
    \pi(\chi_i) = \chi_{\max}^{\alpha / n} \frac{3 \chi_i^2}{\chi_{\max}^3},
\end{equation}
where $n$ is the number of Cepheid host galaxies. This distribution is no longer normalised and alters the prior on $r$ in a strange way. Since $\chi_{\max} = r_{\max} h$, this introduces a spurious factor of $\left(r_{\max} h\right)^{\alpha / n}$ per host galaxy. Thus, the $\max(\bm{\chi})^\alpha$ term implicitly imposes a prior on $H_0$ proportional to $H_0^\alpha$, which~\citetalias{Kenworthy_2022} set to $H_0^{60}$. The simulations from which they ``derive'' this prior include a selection in the observed redshift of $z<0.011$. We have shown that without modelling peculiar velocities or the inhomogeneous Malmquist bias, the correct method for accounting for a redshift selection introduces a factor of $H_0^3$ per host galaxy, as discussed in~\cref{sec:redshift_selection}. Thus, for a uniform-in-volume prior and a redshift-selected sample of 35 host galaxies, the correct treatment of the selection introduces a factor of $H_0^{105}$, rather than $H_0^{60}$.

Adopting~\citetalias{Carrick_2015} as the peculiar velocity model,~\citetalias{Kenworthy_2022} found $H_0 = 76.4^{+2.6}_{-2.4}$ or $74.1^{+2.4}_{-2.1}\kmsecMpc$, depending on whether the redshift scatter $\sigma_v$ is fixed or inferred. In contrast, with the same field and inferring $\sigma_v$ we find $H_0 = 73.8 \pm 1.7\kmsecMpc$. Although in principle our selection treatment effectively introduces a factor of $H_0^{105}$ rather than $H_0^{60}$, direct comparison is complicated by our explicit treatment of both peculiar velocities and inhomogeneous Malmquist. While~\citetalias{Kenworthy_2022} introduced an ad hoc prior modification in an attempt to minimise bias in redshift-limited mocks (which is ultimately incorrect), we present a principled method for modelling redshift selection, as well as properly accounting for the effects of peculiar velocities and inhomogeneous Malmquist bias on it. We demonstrate that our method is unbiased in~\cref{sec:mock_data_bias_tests}.

\subsubsection{Distance selection}

The second case considered by~\citetalias{Kenworthy_2022} is a distance-limited selection, which is presumably meant to model a selection on the observable \ac{SN} magnitudes. In this model, they set the prior on the host galaxy distances to
\begin{equation}\label{eq:kenworthy_dist_prior}
    \pi(\chi_i | d_T) = \frac{f(\chi_i, d_T)}{\int_0^{80} f(x, d_T) \dd x},
\end{equation}
where $d_T$ is the distance cutoff in $h^{-1}\,\mathrm{Mpc}$, the dummy variable $x$ is also in $h^{-1}\,\mathrm{Mpc}$ and
\begin{equation}
    f(\chi, d_T) = \chi^2\,\Phi \left(\frac{\mu(\chi) - \mu(d_T)}{0.15}\right).
\end{equation}
(The inhomogeneous Malmquist term is also included but we omit it for brevity here.) $\Phi$ is the \ac{CDF} of a standard Gaussian distribution defined in~\cref{eq:CDF_standard_normal}. However, since the \ac{CDF} has the limits
\begin{equation}
    \Phi(x) \rightarrow 0 \: \: \: \text{as } x \rightarrow -\infty; \: \: \: \:\Phi(x) \rightarrow 1 \: \: \: \text{as } x \rightarrow +\infty,
\end{equation}
this implies that $\chi$ values \emph{below} $d_T$ are cut off. We assume this is a typo and instead interpret $\Phi$ as the survival function, defined as
\begin{equation}
    \mathrm{SF}(x) = 1 - \Phi(x),
\end{equation}
with which we can reproduce the reported results. Using the \ac{CDF} would force $d_T$ to be less than the distance to the nearest host, since the model cannot assign host galaxies to distances where the prior is forced to zero by the \ac{CDF}.

To verify this, we implement the model of~\citetalias{Kenworthy_2022} and their distance selection (described in~\cref{sec:kenworthy_comparison}). Unlike our analysis, this involves predicting some a priori-calibrated distance estimates, rather than deriving them simultaneously with the other model parameters by forward-modelling the Cepheid magnitudes. For this, we use the Cepheid-derived distances from our redshift-independent analysis in~\cref{sec:Cepheid_only_distances} with a uniform-in-distance-modulus prior, matching the data and assumptions they used. When working in $h^{-1}\,\mathrm{Mpc}$ as they do and marginalising over peculiar velocities with the \ac{LCDM} covariance matrix, we find $H_0 = 71.6 \pm 3.2\kmsecMpc$, consistent with their reported value of $71.6 \pm 4.6\kmsecMpc$ as shown in the second row of their Table~2. (Their uncertainty is larger because they over-estimate the \ac{LCDM} peculiar velocity covariance; see the last paragraph of~\cref{sec:LCDM_covariance}.) Conversely, if we use the model of~\citetalias{Kenworthy_2022} in terms of predicting ``observed'' distances derived from a prior analysis, but work consistently in Mpc without applying any ``selection'' modification to the prior, we again find a lower value of $H_0 = 68.4\pm3.0\kmsecMpc$, consistent with the corresponding no-selection control in that two-step distance-summary model.

However, the approach of~\citetalias{Kenworthy_2022} does not provide a principled method for modelling selection in apparent magnitude. In~\cref{sec:SN_magnitude}, we derive the correct treatment: in a Bayesian hierarchical framework, modifying the prior of latent variables is not justified; instead, the appropriate approach is to apply a correction for the unobserved sources, which is independent of the properties of the detected sources.

\vspace{3mm}

\noindent Although~\citetalias{Kenworthy_2022} do not consider the case of no selection, it is interesting to see how it would play out when measuring distances in $h^{-1}\,\mathrm{Mpc}$ units as they do. This would correspond to $\pi(\chi_i) \propto \chi_i^2$, which, if treated as an improper prior without normalisation (as would appear from their Eq.~15), would produce a factor of $h^3$ per galaxy when mapping it to the prior on $r_i$: a factor of $h^2$ derives from $\chi^2$, and $h$ from the Jacobian of the transformation from $\chi$ to $r$. This yields a prior scaling as $H_0^{105}$, which, as we have shown, is effectively the result of redshift selection. We therefore see that if one works in redshift as the independent variable one does not need to explicitly model redshift selection, analogously to how one does not need to explicitly model selection effects for a volume-limited sample when working in physical distances.

While most of the issues we find with~\citetalias{Kenworthy_2022} are specific to that analysis, we note that the issue of distance priors are pervasive to distance-ladder studies. In general, adopting a uniform-in-distance-modulus or uniform-in-distance prior ($r^0$ or $r^{-1}$) puts galaxies at smaller distances relative to the physically motivated uniform-in-volume prior ($r^2$), biasing $H_0$ high. This may be contributing to a community-wide overestimation of $H_0$~\citep{Desmond_2025}, although it is complicated by selection effects.

\subsubsection{Modelling the peculiar velocity covariance}\label{sec:vpec_cov}

We effectively considered three discrete velocity models: \Manticore, the \citetalias{Carrick_2015} field (both with additional variance), and the \ac{LCDM} covariance without a mean field.~\citetalias{Kenworthy_2022} add an additional covariance to \citetalias{Carrick_2015} describing contributions to the velocity field at small enough scales not captured by~\citetalias{Carrick_2015} yet large enough to couple galaxies, i.e.~not simply a diagonal $\sigma_v$ contribution. They find that the inclusion of this covariance term yields a relatively large shift in $H_0$, pushing it higher. To derive this additional covariance, \citetalias{Kenworthy_2022} use the reconstruction of~\cite{Lilow_2021}, which is a linear reconstruction based on the 2MRS galaxy density field~\citep{Huchra_2012} within a $200\Mpch$ radius. We do not adopt this approach: a covariance of unmodelled peculiar velocities cannot be reliably constructed from two discrete fields. However, it can be estimated from \Manticore, which provides self-consistent posterior samples of the large-scale structure, and accounted for directly by marginalising over field realisations, as we do. Moreover, this approach ensures a self-consistent treatment of the peculiar velocity correlations between host galaxies.

\begin{figure}
    \centering
    \includegraphics[width=\columnwidth]{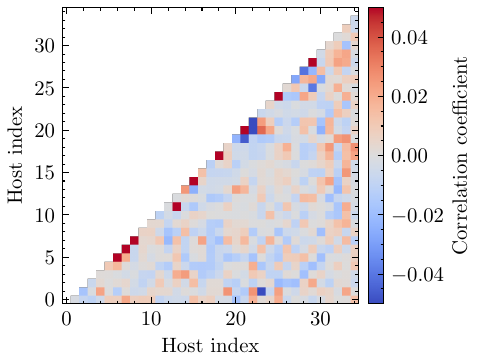}
    \caption{Correlation matrix of the \ac{LCDM} peculiar velocity covariance computed with a non-linear power spectrum including only modes with $k > 0.5\invMpch$, conservatively modelling scales not reconstructed by~\protect\cite{Carrick_2015}. The matrix is shown for the 35 Cepheid host galaxies. Excluding large-scale contributions significantly reduces the correlations between hosts compared to~\protect\cref{fig:host_covariance}, where all scales were included. The resulting correlation coefficients are negligible, indicating that scales below the reconstruction limit of~\protect\cite{Carrick_2015} can be well approximated by a diagonal covariance matrix as in our fiducial model.}
    \label{fig:pecvelcov_correlation_matrix_high_k}
\end{figure}

We also investigate directly whether the covariance between galaxies on scales smaller than those resolved by~\citetalias{Carrick_2015} is significant: we consider the contribution to the \ac{LCDM} peculiar velocity covariance matrix from scales not captured by~\citetalias{Carrick_2015}. The effective smoothing applied to the galaxy density field by~\citetalias{Carrick_2015} is $4\Mpch$, which corresponds to a wavenumber $k\approx 1.6\invMpch$. (\citetalias{Kenworthy_2022} apply an additional Gaussian smoothing of $3\Mpch$.) To be conservative we adopt a cutoff of $k > 0.5\invMpch$ (corresponding to a physical scale of $12.6\Mpch$) and set the power spectrum to zero below this limit when computing the \ac{LCDM} peculiar velocity covariance matrix.~\Cref{fig:pecvelcov_correlation_matrix_high_k} shows the resulting correlation matrix for all 35 Cepheid host galaxies. Compared to~\cref{fig:host_covariance} (computed using the full power spectrum), the correlations are significantly reduced, since it is the large-scale flows that are primarily responsible for coupling the velocities. Of the 595 possible pairs of galaxies, only 7 have an absolute expected \ac{LCDM} peculiar velocity correlation exceeding 0.1, and only 57 exceed 0.02. In comparison, for the full covariance matrix these numbers were 350 and 534, respectively.

To quantify this effect, we infer $H_0$ using the same model as in~\cref{sec:Cepheid_only_H0}, but compare two cases (each assuming \ac{SN} magnitude selection). In the fiducial analysis, when peculiar velocities are accounted for using the~\citetalias{Carrick_2015} reconstruction we assume the covariance matrix to be diagonal with variance $\sigma_v^2$. In the second case, we additionally include the \ac{LCDM} covariance matrix, which accounts for contributions from all scales with $k > 0.5\invMpch$. This fiducial analysis yields $H_0 = 71.4 \pm 1.6$, while the latter yields $71.4\pm1.6\kmsecMpc$. This demonstrates that the off-diagonal contributions are negligible for scales not captured by~\citetalias{Carrick_2015}, justifying our fiducial model where only a diagonal contribution ($\sigma_v$) is added to that. A similar conclusion holds for the other selection models. This disagrees with the finding of~\citetalias{Kenworthy_2022}, which we attribute to them modelling the peculiar velocity covariance from the two distinct fields of~\citetalias{Carrick_2015} and~\cite{Lilow_2021}.

Moreover, we use \Manticore\ which has a spatial resolution of $0.7\Mpch$---hence self-consistently modelling scales smaller than those resolved by~\citetalias{Carrick_2015}---and provides an ensemble of plausible realisations of the local peculiar velocity field describing the reduced cosmic variance given the \ac{BORG} constraints. However, in all cases we find that \Manticore\ and~\citetalias{Carrick_2015} agree to about 1$\sigma$. In~\cite{VF_olympics,McAlpine_2025} we benchmarked \Manticore,~\citetalias{Carrick_2015} and other reconstructions against direct distance data, finding \Manticore\ to be the most accurate model of the local Universe to date, with the~\citetalias{Carrick_2015} reconstruction also performing well. Compared to~\citetalias{Carrick_2015}, the primary advantages of \Manticore\ are its more accurate gravity and galaxy bias models which allow the modelling of the density and velocity field at a higher resolution by going beyond linear theory, and its independent posterior samples of the large-scale structure, over which we marginalise.

The higher fidelity of the \Manticore\ reconstruction is particularly important for host galaxies located within or near galaxy clusters. In~\cite{VF_olympics}, we showed that \ac{BORG}-based models provide significantly improved modelling of galaxies in over-dense regions compared to linear modelling. This is particularly relevant here because at such low redshifts peculiar velocities are the dominant source of uncertainty, and the local dynamics is dominated by infall to nearby clusters, most notably Virgo, Fornax, and Leo. Indeed at least seven of the 35 host galaxies lie within the infall region of these clusters (this number is approximate as we do not carefully test for cluster membership). In contrast,~\citetalias{Carrick_2015} provides only a single mean field estimate.

\subsection{Ramifications of the results and future work}\label{sec:ramifications}

We have shown that a Cepheid-only distance ladder yields $H_0$ values consistent with SH0ES~\citep{Riess_2022}, though using the state-of-the-art \Manticore\ reconstruction to model galaxy bias and account for peculiar velocities gives somewhat lower values unless the sample is assumed to be redshift-selected.
We adopt \Manticore\ as our fiducial peculiar velocity model and compare the inferred $H_0$ with results obtained using the~\citetalias{Carrick_2015} reconstruction.
We find good agreement between the two, though \Manticore\ yields slightly lower values of $H_0$ (see~\cref{fig:H0_stacked,tab:H0_PV_vs_selection} for a summary of all $H_0$ variations).
We picked the \Manticore\ reconstruction because of its validated peculiar velocity field and improved performance relative to other available reconstructions~\citep{VF_olympics,McAlpine_2025}. Additionally, \Manticore\ accurately predicts both the masses and positions of all nearby clusters across all validation metrics (see~\citealt{McAlpine_2025}). We do not include any \acp{SN} in our analysis (except for modelling a \ac{SN} selection), instead focusing on a detailed investigation of the two lower rungs of the distance ladder.

Using \Manticore\ to account for peculiar velocities (and their covariance), the uncertainty on the inferred $H_0$ with only 35 Cepheid host galaxies ranges between $1.2$ and $1.4\kmsecMpc$, which is less than two per cent. Relative to the~\citetalias{Carrick_2015} model, \Manticore\ yields an 18 per cent reduction in the uncertainty on $H_0$ in the fiducial \ac{SN} magnitude-selection case. The relatively small volume probed by the second-rung objects underscores the need for increasingly faithful representations of the local Universe such as those enabled by the \ac{BORG} programme~\citep{Jasche_2013,Jasche_2015,Lavaux_2016,Leclercq_2017,Lavaux_2019,Jasche_2019,Porqueres_2019,Stopyra_2023,McAlpine_2025} to suppress both the statistical and systematic uncertainty due to peculiar velocities. However, assuming these (and other) systematic uncertainties are already under control, further crucial evidence concerning the Hubble tension could be obtained by incorporating more second-rung data.

Even restricting to Cepheids, the host galaxies we used represent only a small subset of available Cepheid measurements. We have used the SH0ES sample because of its high-quality~\citep[e.g.][]{Riess_2022B,Riess_2023,Breuval_2023,Bhardwaj_2023,Riess_2024,Breuval_2024} and self-consistent~\citep{Najeira_2025} measurements, including well-characterised covariances. However, in principle more Cepheid hosts could be incorporated. Beyond that, several alternative second-rung indicators exist, including the \acl{TRGB} (\ac{TRGB};~\citealt{Freedman_2019,Siyang_2024}), Type II \acp{SN}~\citep{Vogl_2024}, J-region asymptotic giant branch (JAGB;~\citealt{jagb}), Mira variables~\citep{mira}, \acl{SBF} (\ac{SBF};~\citealt{Cantiello_2023,Jensen_2025}) and potentially other less explored methods~\citep{Najeira_2025}. \cite{Blakeslee_2021} used an \ac{SBF} calibration anchored to both Cepheid and \ac{TRGB} distances to infer $H_0 = 73.3 \pm 0.7\,\text{(stat)} \pm 2.4\,\text{(sys)}\kmsecMpc$. This measurement has recently been updated with \ac{HST} calibration by~\cite{Jensen_2025}, who found $H_0 = 73.8 \pm 0.7\,\text{(stat)} \pm 2.3\,\text{(sys)}\kmsecMpc$. The dominant systematic error arises primarily from the distance calibration, suggesting the value of a principled joint calibration approach similar to the one we undertake here. A future analysis should revisit this kind of approach to model jointly the Cepheid, \ac{TRGB}, and \ac{SBF} (and perhaps other) distances using the framework outlined here.

It is likely that with additional distance data and ever-improving peculiar velocity modelling (such as with \Manticore), second-rung inferences could become even more precise, with the key added benefit of being independent of \acp{SN} and associated potential systematics. This is even before considering future data:~\citet{HWO} argue that only 24 new Cepheid hosts observed by the proposed Habitable Worlds Observatory (HWO) would be sufficient to reach a one per cent determination of $H_0$ using the SH0ES error model. As of now, with \Manticore\ we achieve a determination of $H_0$ to better than two per cent (up to the caveat of uncertain selection).

A two-rung ladder is particularly useful because of its complete elimination of \acp{SN}, beyond their use to select nearby hosts. Indeed there have been several studies suggesting possible miscalibration of \acp{SN} (e.g.~\citealt{Seifert,Efs_SNe,SN1,SN4,Son}). As an example, a recent exploration by~\cite{Wojtak_2025} modelled the SH0ES \ac{SN} sample as comprising two populations with distinct intrinsic and extrinsic properties---primarily differences in mean absolute magnitude and extinction coefficients---finding a preference for lower $H_0$ values and reducing the Hubble tension by 30 to 50 per cent. Indeed, an alternative programme to SH0ES---the Chicago-Carnegie Hubble Program (CCHP;~\citealt{Freedman_2019})---infers $H_0 = 69.8 \pm 0.8~(\text{stat}) \pm 1.7 ~(\text{sys}) \kmsecMpc$ using the \ac{TRGB}. This was recently refined using James Webb Space Telescope data of \ac{TRGB}, Cepheids and the J-region asymptotic giant branch, finding very similar results~\citep{Freedman_2025}.

\section{Conclusion}\label{sec:conclusion}

The persistence of the Hubble tension demands that we multiply cross-check the cosmic distance ladder to vet the local $H_0$ measurement fully. To this end, we have considered dropping the \acp{SN} from the SH0ES pipeline to constrain $H_0$ from the Cepheid distances (with geometric anchors) and host galaxy redshifts alone. To minimize the risk of introducing new systematics, we keep our data as close as possible to the SH0ES pipeline~\citep{Riess_2022}. This follows in part the analysis of~\citetalias{Kenworthy_2022}, who find $H_0=72.9^{+2.4}_{-2.2}\kmsecMpc$, where the uncertainty includes both statistics and systematics. We find several issues with this analysis, mostly to do with selection modelling. We introduce a principled framework for modelling selection effects and also employ state-of-the-art peculiar velocity modelling with \Manticore~\citep{McAlpine_2025}, including marginalisation over plausible realisations of the local Universe. We summarise the inferred $H_0$ in~\cref{fig:H0_stacked}. Our main findings are as follows:
\begin{enumerate}
	\item Our fiducial model, which accounts for peculiar velocities using~\Manticore, yields $71.1 \pm 1.4\kmsecMpc$ when the Cepheid host sample is assumed to be selected on the basis of \ac{SN} apparent magnitudes, which we consider the more plausible scenario. If however the sample were selected on observed redshift, we would infer $H_0 = 72.5 \pm 1.4\kmsecMpc$. These two cases bracket the range of plausible $H_0$ values, with mixed selection scenarios lying in between. Our headline result is in $2.8\sigma$ tension with combined \ac{CMB} data from ACT, SPT and \textit{Planck}~\citep{Camphuis_2025}.
	    \item While~\cite{VF_olympics,McAlpine_2025} have demonstrated that \Manticore\ provides the most accurate velocity field among all currently in the literature, we explore others to investigate possible systematic uncertainties. Using the~\citetalias{Carrick_2015} reconstruction shifts the inferred $H_0$ marginally higher, as well as increasing the uncertainty on $H_0$ by 22 per cent in the fiducial selection case. Any $H_0$ inference with second-rung objects only is limited by the peculiar velocity modelling, highlighting the need for a state-of-the-art reconstruction like \Manticore.
    \item We identify several problems in the~\citetalias{Kenworthy_2022} analysis of the same data: \emph{i)} they adopt an unphysical prior on the galaxy distances used as input to their analysis ($r^{-1}$ rather than the uniform-in-volume $r^2$; see also~\citealt{Desmond_2025}), \emph{ii)} they model selection effects as acting on the prior of the latent true distances, rather than modelling the unobserved sources, and \emph{iii)} by representing distances in $h^{-1}\,\mathrm{Mpc}$ rather than Mpc units (i.e. actually redshifts), they introduce spurious high powers of $H_0$ in the prior.
\end{enumerate}

We conclude that the SH0ES Cepheid sample alone yields a Hubble constant consistent with the full three-rung SH0ES analysis and with only slightly larger uncertainty, though with a $\sim$1.3$\sigma$ lower central value in the fiducial magnitude-selection scenario.
This conclusion relies on careful modelling of both the selection function and peculiar velocities using the state-of-the-art \Manticore\ model. The ongoing challenge of understanding and resolving the Hubble tension motivates further work on both the second rung of the distance ladder, independent of \acp{SN}, and on the modelling of \acp{SN} themselves (e.g.~\texttt{BayeSN};~\citealt{Mandel_2022,Grayling_2024}).

More constructively, our analysis demonstrates that $H_0$ may be inferred with high precision (less than two per cent statistical uncertainty) from first-plus-second-rung data alone, extending only to $z\approx0.01$. This suggests that bringing further such data to bear may readily afford a Hubble constant inference even more precise than SH0ES and crucially avoiding the third rung (e.g.~\acp{SN}) altogether. We stress that while the second-rung is independent of \acp{SN} (except when modelling the host selection), it is susceptible to significant systematics arising from selection effects or peculiar velocity modelling if these are not properly accounted for. Thus, this endeavour requires principled methods for accounting for selection (as we have implemented here) as well as accurate density and velocity field models of the local Universe, which are growing in quality through Bayesian forward modelling of the galaxy number density field (\ac{BORG}). Finally, we remark that the sensitivity to selection effects underscores the need for observing programmes to adopt well-defined selection procedures that render the required cosmological inference procedure unambiguous.

\section{Data availability}

The SH0ES and \PP\,data is available at \url{https://github.com/PantheonPlusSH0ES/DataRelease}. The \Manticore\ resimulations will be made publicly available at \url{https://www.cosmictwin.org}. The~\cite{Carrick_2015} velocity reconstruction is available at \url{https://cosmicflows.iap.fr}. The code and all other data will be made available on reasonable request to the authors. The code will also be released publicly on publication of the paper.

\section*{Acknowledgements}

We thank Indranil Banik, Karim Benabed, Nicholas Choustikov, Julien Devriendt, George Efstathiou, Pedro Ferreira, Sebastian von Hausegger, Mike Hudson, D'Arcy Kenworthy, Florent Leclercq, Gabriele Montefalcone, Jos\'e Antonio N\'ajera, John Peacock, Erik Peterson, Adam Riess, Subir Sarkar, Dan Scolnic, Ian Steer, Licia Verde and Tariq Yasin for useful inputs and discussions. This work was done within the Aquila Consortium.\footnote{\url{https://aquila-consortium.org}}

RS acknowledges financial support from STFC Grant No. ST/X508664/1, the Snell Exhibition of Balliol College, Oxford, and the CCA Pre-doctoral Program. HD is supported by a Royal Society University Research Fellowship (grant no. 211046). ET was supported by STFC through Imperial College Astrophysics Consolidated Grant ST/W000989/1. JJ, GL and SM acknowledge support from the Simons Foundation through the Simons Collaboration on ``Learning the Universe''. JJ and SM acknowledge support by the research project grant ``Understanding the Dynamic Universe,'' funded by the Knut and Alice Wallenberg Foundation (Dnr KAW 2018.0067). Additionally, JJ acknowledges financial support from the Swedish Research Council (VR) through the project "Deciphering the Dynamics of Cosmic Structure" (2020-05143) and GL acknowledges support from the CNRS IEA programme ``Manticore''.

The authors would like to acknowledge the use of the University of Oxford Advanced Research Computing (ARC) facility in carrying out this work\footnote{\url{https://doi.org/10.5281/zenodo.22558}}. In addition, this work has made use of the Infinity Cluster hosted by Institut d'Astrophysique de Paris, and was granted access to the HPC resources of TGCC (Très Grand Centre de Calcul), Irene-Joliot-Curie supercomputer, under the allocations A0170415682 and SS010415380.

\bibliographystyle{mnras}
\bibliography{ref}

\appendix

\section{Mock data bias tests}\label{sec:mock_data_bias_tests}

We assess the impact of selection on either \ac{SN} apparent magnitude or host galaxy redshift using mock catalogues. We generate mock catalogues designed to match the properties of the observed data and demonstrate that the inferred value of $H_0$ remains unbiased.

The \ac{SN} magnitude-selected sample is generated as follows. For simplicity, rather than modelling the full \ac{CPLR}, we assume all Cepheid stars have identical absolute magnitudes. We draw source galaxy distances from a uniform-in-volume ($r^2$) distribution. For each trial, we compute the \ac{SN} apparent magnitude, given an assumed $M_B = -19.25$, adding a scatter of $\sigma_{\rm SN} = 0.15~\mathrm{dex}$. If the resulting apparent magnitude is below the selection threshold, we retain the sampled distance; otherwise, we repeat the process. For each accepted host galaxy, we generate five Cepheids (the precise number is unimportant), each at the same distance, with absolute magnitude $\tilde{M}_W = -18$ and a scatter in observed magnitude of $\sigma_{\rm Ceph} = 0.1~\mathrm{dex}$. The observed redshift of the host galaxy is then computed, assuming a given $H_0 = 70\kmsecMpc$ and including a velocity dispersion of $\sigma_v = 250\kmsec$ to mimic peculiar velocities. This procedure is repeated until a sample of 35 host galaxies is obtained. Finally, to mimic the inclusion of the \ac{LMC} and NGC\,4258 as calibrators, we add two nearby sources. For each, we assign a true distance modulus and sample an ``observed'' distance modulus with a scatter of $0.25~\mathrm{dex}$, similar to the values reported in the literature. The host galaxy redshift-limited sample is generated analogously, except that \ac{SN} apparent magnitudes are not modelled; instead, the rejection process is based on whether the host galaxy redshift falls below a specified threshold. When sources are drawn from an $r^2$ distribution, a maximum distance must be imposed for practical reasons. We adopt an upper limit of 150 Mpc, which ensures that all sources passing either the magnitude or redshift selection criteria lie well below this threshold (typically within 50 Mpc). We also test an alternative procedure in which we generate a fixed total number of galaxies and then retain those that pass the cut, rather than resampling until a fixed number pass the cut. In both cases our procedure is unbiased.

\begin{figure*}
    \centering
    \includegraphics[width=\textwidth]{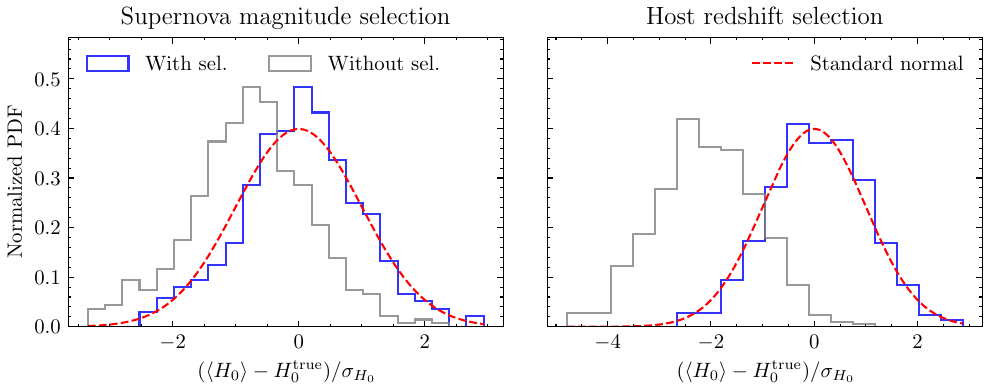}
    \caption{Bias test of $H_0$ for ``Cepheid''-only mock inferences, as described in~\cref{sec:mock_data_bias_tests}, under host selection by either \ac{SN} magnitude or host redshift. When selection effects are not modelled (grey), the inferred $H_0$ is significantly biased low. By contrast, when selection is properly modelled (blue), the inference is unbiased in both cases, as indicated by excellent agreement with the standard normal distribution (red dashed).}
    \label{fig:H0_bias_mock}
\end{figure*}

In~\cref{fig:H0_bias_mock}, we present bias tests in $H_0$ for the \ac{SN} magnitude- and redshift-limited selection. In each case, we use identical injected parameter values but draw the mock catalogue using different random seeds. This procedure is repeated 500 times. The bias is defined as
\begin{equation}
    \mathrm{Bias}(H_0) = \frac{\langle H_0 \rangle - H_0^{\rm true}}{\sigma_{H_0}},
\end{equation}
where $\langle H_0 \rangle$ and $\sigma_{H_0}$ denote the posterior mean and standard deviation of the inferred $H_0$, and $H_0^{\rm true}$ is the true (injected) value. If the inferred $H_0$ is unbiased (and its posterior is Gaussian), then the distribution of $\mathrm{Bias}(H_0)$ across the mock realisations should follow a standard normal distribution.~\cref{fig:H0_bias_mock} demonstrates that, when selection effects are properly modelled, the recovered $H_0$ is unbiased on average in all cases. In contrast, neglecting selection effects introduces a significant bias in the inferred $H_0$ for each scenario.

\section{Local Universe reconstructions}\label{sec:universe_reconstructions}

As discussed in~\cref{sec:methodology}, some of our peculiar velocity models use the density and velocity field reconstruction of either~\Manticore\ or~\citetalias{Carrick_2015}. This enables us to both model the inhomogeneous Malmquist bias as a function of the local density, as well as the contribution of the \ac{LOS} peculiar velocities to observed redshifts. Both reconstructions are based on the \TWOMPP\ galaxy catalogue~\citep{Lavaux_2011}, which contains \num{69160} galaxies and combines 2MASS photometry~\citep{Skrutskie_2006} with redshifts from 2MRS~\citep{Huchra_2012}, 6dF~\citep{Jones_2009}, and \ac{SDSS} DR7~\citep{Abazajian_2009}. Apparent magnitudes are corrected for Galactic extinction, $k$-corrections, evolution, and surface brightness dimming. The \TWOMPP\ sample is magnitude-limited to $K < 11.5$ in the 2MRS region and $K < 12.5$ in the 6dF and \ac{SDSS} regions.

The density and velocity field of either \Manticore\ or~\citetalias{Carrick_2015} is reconstructed in real space. Since the real-space distances to the host galaxies are not known a priori, we consider all plausible velocities along the \ac{LOS} to each source. The predicted peculiar velocity of the $i$\textsuperscript{th} host is given by
\begin{equation}
    V_{\mathrm{pec},i}^{\rm C15} = \left(\beta \bm{V}(\bm{r}_i) + \Vext\right) \cdot \hat{\bm{r}}_i,
\end{equation}
where $\bm{V}(\bm{r}_i)$ is the velocity field evaluated at candidate galaxy position $\bm{r_i}$. $\beta$ is set to unity for \Manticore\ and treated as a free parameter when using the reconstruction of~\citetalias{Carrick_2015} (where it is related to $\sigma_8$). The associated density field can be used to model galaxy bias $n_g(\bm{r})$, i.e.~the so-called inhomogeneous Malmquist bias, which is the prior preference for galaxies to be located in regions of higher matter density. This leads to the following prior on the 3D position
\begin{equation}\label{eq:host_prior_IM}
    \pi(\bm{r} \mid \bm{\theta}) \propto n_g(\bm{r},\,\bm{\theta}).
\end{equation}
The omitted normalisation is the volume integral of $n_g(\bm{r},\,\bm{\theta})$.
The same volume normalisation $\int n_g\,\dd^3\bm{r}$ enters both the host distance priors and the selection probability of~\cref{eq:selection_normalisation_SN}, and the two cancel exactly in the posterior.
Both reconstructions store the density (and velocity) field on a grid in \Mpch.
We therefore convert each sampled position from physical Mpc to \Mpch\ using the sampled value of $H_0$ before querying $n_g$.
We discuss the galaxy bias choices for \Manticore\ and~\citetalias{Carrick_2015} below.

\subsection{The \texttt{Manticore-Local} reconstruction}\label{sec:Manticore_reconstruction}

\texttt{Manticore} is the latest set of reconstructions using the \ac{BORG} algorithm~\citep{Jasche_2013,Jasche_2015,Lavaux_2016,Leclercq_2017,Jasche_2019,Lavaux_2019,Porqueres_2019,Stopyra_2023,McAlpine_2025}, which infers a posterior distribution of voxel-by-voxel densities at $z=1000$ (and hence at all lower redshifts) by employing a gravity forward model, while accounting for redshift-space distortions, observational selection effects and a galaxy bias model, and comparing the resulting predictions to the observed galaxy number density field with a Poisson likelihood. In particular we use the \Manticore\ run, which constrains the initial conditions to match the matter distribution of the local Universe applied to the \TWOMPP\ data.

\begin{figure*}
    \centering
    \includegraphics[width=\textwidth]{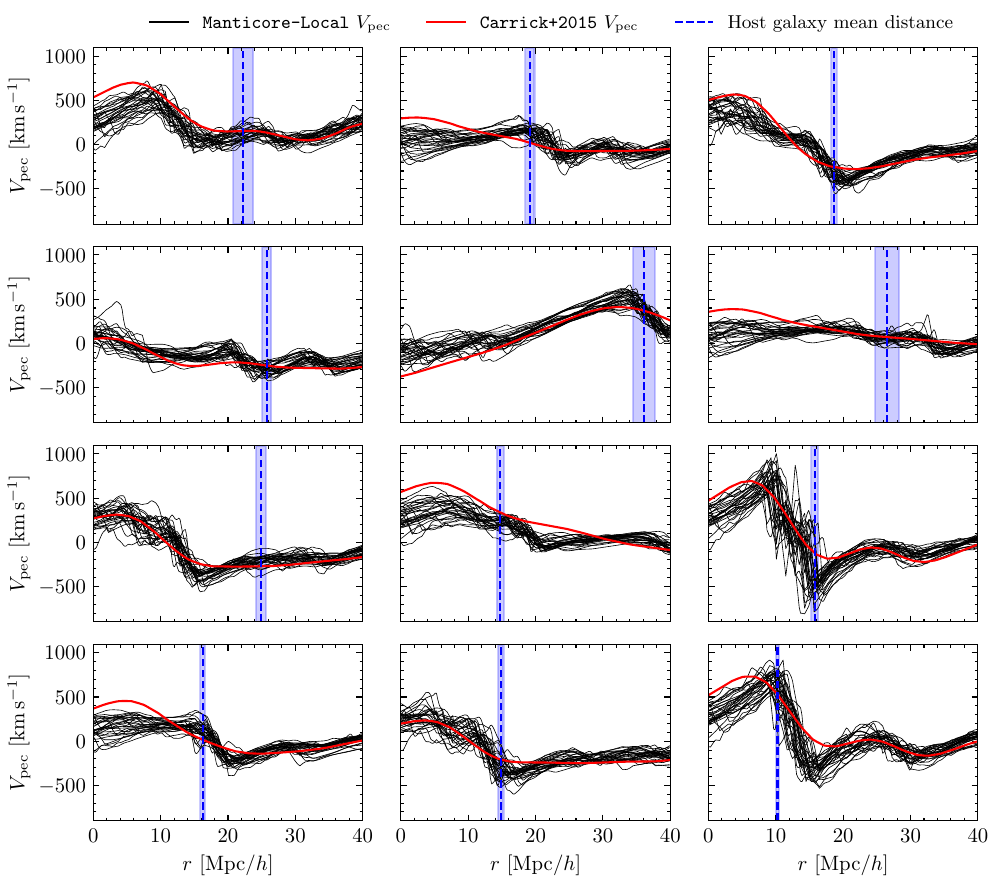}
    \caption{Comparison of radial peculiar velocities, $\Vpec$, between \Manticore\ (black) and~\protect\cite{Carrick_2015} (red) along the \ac{LOS} to 12 randomly selected Cepheid host galaxies, whose mean inferred distances are shown by the blue vertical lines with shaded bands indicating $1\sigma$ uncertainty. Each panel shows 30 \Manticore\ realisations, representing independent draws from the \ac{BORG} posterior, while~\protect\cite{Carrick_2015} provides only a mean field estimate. The two reconstructions are generally in good agreement. In several cases, the host galaxies reside in clusters, as indicated by the characteristic infall signature in $\Vpec$.}
    \label{fig:LOS_Vpec}
\end{figure*}

We use 80 \Manticore\ realisations~\citep{McAlpine_2025}.
These were simulated with the same COLA gravity model~\citep{Tassev_2013} and numerical settings used during the \ac{BORG} inference.
Each realisation evolves independent posterior initial conditions inferred on a $256^3$ grid, corresponding to a spatial resolution of $2.65~\Mpch$.
The density and velocity fields are computed from the final particle distribution using the \ac{PCS} mass-assignment scheme and rendered on the same $256^3$ grid.
In~\cref{fig:LOS_Vpec} we compare the \ac{LOS} velocities for a sample of Cepheid host galaxies between the \Manticore\ and~\citetalias{Carrick_2015} reconstructions, finding very good agreement.

\Manticore\ assumes cosmological parameters from the Dark Energy Survey Year 3 (DES Y3;~\citealt{DES_Y3}) ``3$\times$2pt + All Ext.'' $\Lambda$CDM cosmology: $h = 0.681$, $\Omega_{\mathrm{m}} = 0.306$, $\Omega_{\Lambda} = 0.694$, $\Omega_{\mathrm{b}} = 0.0486$, $A_\mathrm{s} = 2.099 \times 10^{-9}$, $n_\mathrm{s} = 0.967$, and $\sigma_8 = 0.807$. Thus, unlike~\citetalias{Carrick_2015}, which is derived independently of the growth factor and consequently of $\sigma_8$, these are assumed in \Manticore. However, by introducing a free parameter $\beta$ (see~\cref{sec:Carrick_reconstruction}) to scale the predicted peculiar velocities, we can (approximately) assess the impact of varying $\sigma_8$---we find it to have no impact on the inferred $H_0$.

Moreover, while \Manticore\ assumes $H_0 = 68.1\kmsecMpc$, we assume our inference to be insensitive to using it while varying $H_0$. This is because the deviations we probe from the fiducial $H_0$ are small and because the hosts typically reside in relatively low-density regions (see \citealt{Tsaprazi_2025} and Table~2 of~\citealt{Tsaprazi_2022}). Moreover, when querying \Manticore\ for either density or velocity, we consistently convert sampled distances from $\mathrm{Mpc}$ to $h^{-1}\,\mathrm{Mpc}$ using the sampled value of $H_0$. This removes the first-order effect of varying $H_0$.

To model the galaxy bias $n_g(r)$ we adopt a phenomenological smooth double power-law,
\begin{equation}
\begin{aligned}
    n_g(r,\,\bm{\theta}) &=
    \left(\frac{\rho(r)}{\rho_t}\right)^{\alpha_{\rm low}} \\
    &\quad \times
    \left[
        1 + \left(\frac{\rho(r)}{\rho_t}\right)^{1 / \Delta_{\ln\rho}}
    \right]^{(\alpha_{\rm high} - \alpha_{\rm low})\Delta_{\ln\rho}},
\end{aligned}
\end{equation}
where $\rho(r)$ is the density at distance $r$ along the \ac{LOS} to the host, $\alpha_{\rm low}$ and $\alpha_{\rm high}$ are the two slopes, and $\rho_t$ is the transition density, all of which we infer jointly with the other parameters.
The parameter $\Delta_{\ln\rho}$ controls the transition width in $\ln\rho$.
We place a truncated Gaussian prior on $\alpha_{\rm low}$, centred at 1 with unit standard deviation and restricted to positive values, and on $\alpha_{\rm high}$, centred at 0.5 with unit standard deviation and likewise truncated at zero. For $\ln \rho_t$ we adopt a zero-centred Gaussian prior with standard deviation of two. For $\Delta_{\ln\rho}$ we adopt a Gaussian prior with mean $0.5$ and standard deviation $0.5$, truncated to $0.05 \le \Delta_{\ln\rho} \le 3$.

We also consider a model in which the small-scale velocity dispersion depends on local density to capture higher unresolved dispersion in clusters,
\begin{equation}\label{eq:sigma_v_density}
    \sigma_v(\delta)
    =
    \sigma_{v,\mathrm{low}} + \frac{\sigma_{v,\mathrm{high}} - \sigma_{v,\mathrm{low}}}{1 + \left(\frac{1 + \delta}{\rho_{\sigma_v}}\right)^{-k_{\sigma_v}}},
\end{equation}
where $\delta$ is the matter overdensity, $\sigma_{v,\mathrm{low}}$ and $\sigma_{v,\mathrm{high}}$ set the asymptotic dispersions in low and high density regions, $\rho_{\sigma_v}$ is the transition density, and $k_{\sigma_v}$ controls the steepness of the sigmoid. We adopt the following priors: $\sigma_{v,\mathrm{low}}$ and $\sigma_{v,\mathrm{high}}$ each follow a Maxwell distribution with scale $200\kmsec$ (mean of approximately $320\kmsec$), $\ln \rho_{\sigma_v}$ follows a normal distribution with mean of unity and standard deviation of five, and $k_{\sigma_v}$ follows a truncated normal with mean of unity, standard deviation of unity, and lower bound at zero. Our fiducial analysis retains the simpler assumption of a constant $\sigma_v$, but Table~\ref{tab:H0_PV_vs_selection} shows that this density-dependent model yields almost identical results.

Instead of using a single field, for \Manticore\ we employ a set of 30 posterior samples (30 independent density and velocity fields from $N$-body simulations with constrained initial conditions) over which we marginalise. Following~\cite{VF_olympics}, we do this by averaging the likelihood over the \Manticore\ posterior samples.

\subsection{The Carrick et al.~2015 reconstruction}\label{sec:Carrick_reconstruction}

\citetalias{Carrick_2015} reconstructs the luminosity-weighted density field from redshift-space galaxy positions in the \TWOMPP\ catalogue~\citep{Lavaux_2011} using the iterative scheme of~\citet{Yahil_1991}. The velocity field is derived from the density field using linear theory and scaled by a free parameter $\beta$, defined as
\begin{equation}\label{eq:beta_star}
    \beta \equiv \frac{f \sigma_{8,\mathrm{NL}}}{\sigma_{8,g}},
\end{equation}
where $f \approx \Om^{0.55}$ is the dimensionless growth rate in \ac{LCDM}~\citep{Bouchet_1995,Wang_1998}. Here, $\sigma_{8,g}$ denotes the fluctuation amplitude in the galaxy field, while $\sigma_{8,\mathrm{NL}}$ corresponds to the non-linear matter field. In the \TWOMPP\ data, $\sigma_{8,g}$ is estimated to be $0.98 \pm 0.07$~\citep{westover} or $0.99 \pm 0.04$~\citep{Carrick_2015}. The velocity field is computed on a $256^3$ grid within a $400\Mpch$ box, assuming $\Om = 0.3$. This reconstruction has been used to constrain structure growth and the $S_8$ parameter using peculiar velocity data~\citep[e.g.][]{Boruah_2019,Said_2020,Boubel_2024,VF_olympics}, and is a standard tool to correct for peculiar velocities in $H_0$ inferences~\citep[e.g.][]{Guidorzi_2017,Boruah_2021,Peterson_2022,Brout_2022,Boubel_2024B,Boruah_2025}. Since the~\citetalias{Carrick_2015} reconstruction is linear, it is appropriate to adopt a simple linear bias model of the form
\begin{equation}
    n_g(r) = 1 + b_1 \delta(r),
\end{equation}
where $b_1$ is the linear bias parameter. To maintain consistency with the earlier definition of $\beta$, we set $b_1 \equiv f / \beta$.

\section{Theory peculiar velocity covariance}\label{sec:LCDM_covariance}

The peculiar velocity covariance matrix quantifies correlations in \ac{LOS} velocities induced by large-scale structure~\citep{Kaiser_1998,Hui_2006,Davis_2011}. For galaxies indexed by $i$ and $j$, it is defined as
\begin{equation}
    \xi_{ij} = \left\langle (\bm{v}_i \cdot \hat{\bm{r}}_i)(\bm{v}_j \cdot \hat{\bm{r}}_j) \right\rangle,
\end{equation}
and is computed from the \ac{LCDM} matter power spectrum as
\begin{equation}\label{eq:LCDM_cov}
    \xi_{ij} = \frac{\dd D_i}{\dd\tau} \frac{\dd D_j}{\dd\tau} \int \frac{\dd k}{2\pi^2} P(k) F(k r_i,\,k r_j,\,\hat{\bm{r}}_i \cdot \hat{\bm{r}}_j),
\end{equation}
where $D_i$ is the linear growth factor at the redshift of the $i$\textsuperscript{th} galaxy, and $\tau$ is the conformal time with $\dd\tau \equiv \dd t / a$.
$F$ is defined by
\begin{equation}\label{eq:LCDM_F}
    F(u, v, \cos\theta) = \sum_{\ell = 0}^{\infty} (2\ell + 1) j^\prime_\ell(u) j^\prime_\ell(v) P_\ell(\cos \theta)
\end{equation}
where $j^\prime_\ell(x)$ is the derivative of the spherical Bessel function of order $\ell$ with respect to its argument and $P_\ell(x)$ is the Legendre polynomial of order $\ell$. We evaluate~\cref{eq:LCDM_cov} using the non-linear matter power spectrum $P(k)$ from \texttt{CAMB}~\citep{CAMB} over the interval $k \in [10^{-4},\,20]~h / \mathrm{Mpc}$, assuming the fiducial \textit{Planck} cosmology~\citep{Planck_2020_cosmo}. We verify that changing the background cosmology, including variations in the assumed value of $H_0$ when computing $P(k)$ (e.g.~switching from the \textit{Planck} to the SH0ES value), does not affect the inferred value of $H_0$. We truncate the sum in~\cref{eq:LCDM_F} at $\ell = 2000$, which is sufficient for convergence.

For the special case of considering the same object so that $\cos\theta = 1$ and $r_i = r_j = r$, we have that $F(k r,\,k r,\, 1) = 1/3$, yielding the diagonal elements
\begin{equation}
    \xi_{ii} = \left( \frac{\dd D_i}{\dd \tau} \right)^2 \int \frac{\dd k}{6\pi^2} P(k).
\end{equation}
We find a value of approximately $255\kmsec$ for our choice of the non-linear matter power spectrum $P(k)$. Adopting a linear $P(k)$ instead yields diagonal elements of approximately $200\kmsec$. This also contrasts with~\citetalias{Kenworthy_2022}, who incorrectly claim that the diagonal velocity dispersion is approximately $380\kmsec$ when computed using a non-linear matter power spectrum, leading to an overestimation of their uncertainty. We have managed to reproduce this number by using inconsistent units for $P(k)$ and $k$.

\section{Cepheid host properties}\label{sec:host_table}

In~\cref{tab:cepheid_hosts} we list the Cepheid host galaxy Galactic coordinates, \ac{CMB}-frame redshifts, distance moduli, and peculiar velocities as inferred from both \Manticore\ and~\citetalias{Carrick_2015}. For both inferences, we assume \ac{SN} magnitude selection. Comparing their peculiar velocities, we find the two models to be generally consistent. The \Manticore\ peculiar velocities assume $\beta = 1$, include the sampled $\Vext$, and are computed by stacking the velocities from each \ac{MCMC} step and across the 80 \Manticore\ realisations. In contrast, the peculiar velocities from~\citetalias{Carrick_2015} are computed from the mean reconstructed velocity field while jointly sampling $\beta$ and $\Vext$; their quoted uncertainties therefore represent only the posterior width in those parameters, not of the large-scale structure reconstruction uncertainty. As a comparison, the \Manticore\ model yields a residual velocity scatter of $\sigma_v = 172.5 \pm 27.3\kmsec$, whereas the~\citetalias{Carrick_2015} model gives $\sigma_v = 216.6 \pm 32.1\kmsec$.

\begin{table*}
    \centering
    \small
    \begin{tabular}{lcccccc}
        \toprule
        Name & $\ell~[\deg]$ & $b~[\deg]$ & $cz_{\rm CMB}~[\kmsec]$ & $\mu_{\rm host}~[\mathrm{mag}]$ & $V_{\rm pec}^{\Manticore}~[\kmsec]$ & $V_{\rm pec}^{\rm C15}~[\kmsec]$ \\
        \midrule
        M101   & 102 &  60 &  366 & $29.17_{-0.04}^{+0.04}$ & $ -81_{-84}^{+99}$   & $ -14_{-66}^{+73}$ \\
        M1337  & 303 &  53 & 2896 & $32.91_{-0.07}^{+0.07}$ & $ 155_{-73}^{+73}$   & $ 105_{-55}^{+59}$ \\
        N0691  & 141 & -39 & 2581 & $32.76_{-0.08}^{+0.07}$ & $ 147_{-72}^{+82}$   & $ 191_{-67}^{+64}$ \\
        N1015  & 172 & -54 & 2401 & $32.55_{-0.06}^{+0.06}$ & $  29_{-66}^{+76}$   & $  -2_{-57}^{+55}$ \\
        N1309  & 202 & -53 & 2003 & $32.51_{-0.05}^{+0.05}$ & $-180_{-105}^{+99}$  & $-155_{-56}^{+59}$ \\
        N1365  & 238 & -55 & 1379 & $31.36_{-0.05}^{+0.05}$ & $ -17_{-108}^{+150}$ & $ -46_{-62}^{+65}$ \\
        N1448  & 252 & -51 & 1097 & $31.28_{-0.04}^{+0.04}$ & $ -46_{-129}^{+132}$ & $ -50_{-67}^{+68}$ \\
        N1559  & 274 & -41 & 1304 & $31.46_{-0.05}^{+0.05}$ & $ -93_{-94}^{+108}$  & $ -76_{-78}^{+75}$ \\
        N2442  & 281 & -22 & 1544 & $31.46_{-0.06}^{+0.06}$ & $ 166_{-88}^{+80}$   & $ 140_{-82}^{+80}$ \\
        N2525  & 232 &  11 & 1787 & $32.00_{-0.07}^{+0.07}$ & $  25_{-79}^{+81}$   & $ -65_{-78}^{+74}$ \\
        N2608  & 195 &  34 & 2386 & $32.70_{-0.06}^{+0.06}$ & $-191_{-78}^{+78}$   & $-298_{-81}^{+79}$ \\
        N3021  & 192 &  51 & 1775 & $32.37_{-0.09}^{+0.10}$ & $-355_{-91}^{+102}$  & $-394_{-75}^{+78}$ \\
        N3147  & 136 &  39 & 2977 & $33.18_{-0.07}^{+0.07}$ & $-294_{-89}^{+94}$   & $-319_{-78}^{+89}$ \\
        N3254  & 200 &  59 & 1649 & $32.34_{-0.06}^{+0.06}$ & $-430_{-87}^{+83}$   & $-405_{-70}^{+71}$ \\
        N3370  & 225 &  60 & 1619 & $32.13_{-0.05}^{+0.05}$ & $-353_{-97}^{+105}$  & $-329_{-67}^{+64}$ \\
        N3447  & 228 &  61 & 1394 & $31.93_{-0.03}^{+0.04}$ & $-205_{-104}^{+98}$  & $-296_{-68}^{+64}$ \\
        N3583  & 158 &  62 & 2317 & $32.80_{-0.06}^{+0.06}$ & $-313_{-74}^{+68}$   & $-403_{-69}^{+75}$ \\
        N3972  & 139 &  60 & 1106 & $31.69_{-0.07}^{+0.06}$ & $-465_{-99}^{+120}$  & $-364_{-71}^{+79}$ \\
        N3982  & 139 &  60 & 1106 & $31.65_{-0.06}^{+0.06}$ & $-453_{-107}^{+129}$ & $-359_{-70}^{+78}$ \\
        N4038  & 287 &  42 & 1979 & $31.66_{-0.09}^{+0.10}$ & $ 340_{-86}^{+75}$   & $ 350_{-59}^{+66}$ \\
        N4424  & 284 &  71 &  767 & $30.83_{-0.10}^{+0.12}$ & $ 452_{-298}^{+340}$ & $ 401_{-84}^{+82}$ \\
        N4536  & 293 &  65 & 1049 & $30.84_{-0.05}^{+0.05}$ & $ 646_{-243}^{+145}$ & $ 467_{-64}^{+63}$ \\
        N4639  & 294 &  76 & 1385 & $31.79_{-0.10}^{+0.08}$ & $-414_{-125}^{+199}$ & $-182_{-58}^{+57}$ \\
        N4680  & 301 &  51 & 2791 & $32.73_{-0.12}^{+0.11}$ & $ 207_{-79}^{+77}$   & $ 169_{-54}^{+58}$ \\
        N5468  & 335 &  53 & 2992 & $33.10_{-0.05}^{+0.06}$ & $  28_{-68}^{+75}$   & $ -30_{-52}^{+57}$ \\
        N5584  & 345 &  55 & 1904 & $31.81_{-0.05}^{+0.05}$ & $ 142_{-94}^{+99}$   & $  22_{-55}^{+56}$ \\
        N5643  & 321 &  15 & 1433 & $30.52_{-0.05}^{+0.05}$ & $ 351_{-105}^{+91}$  & $ 451_{-79}^{+78}$ \\
        N5728  & 337 &  38 & 3148 & $33.02_{-0.07}^{+0.10}$ & $ 150_{-74}^{+73}$   & $ 121_{-63}^{+69}$ \\
        N5861  & 349 &  39 & 2179 & $32.28_{-0.09}^{+0.08}$ & $ 175_{-97}^{+84}$   & $  51_{-63}^{+70}$ \\
        N5917  & 355 &  40 & 2108 & $32.36_{-0.07}^{+0.07}$ & $   1_{-100}^{+95}$  & $ -55_{-63}^{+70}$ \\
        N7250  &  94 & -14 &  878 & $31.52_{-0.10}^{+0.10}$ & $-375_{-70}^{+73}$   & $-398_{-74}^{+69}$ \\
        N7329  & 321 & -46 & 3124 & $33.26_{-0.07}^{+0.07}$ & $  79_{-89}^{+84}$   & $  11_{-90}^{+79}$ \\
        N7541  &  83 & -51 & 2305 & $32.60_{-0.07}^{+0.07}$ & $ -18_{-94}^{+80}$   & $ -70_{-57}^{+58}$ \\
        N7678  &  99 & -37 & 3145 & $33.23_{-0.07}^{+0.07}$ & $  75_{-74}^{+84}$   & $  88_{-64}^{+57}$ \\
        U9391  & 101 &  53 & 1991 & $32.79_{-0.06}^{+0.06}$ & $-446_{-82}^{+81}$   & $-385_{-71}^{+76}$ \\
        \bottomrule
    \end{tabular}
    \caption{Summary of the 35 Cepheid host galaxies used in our analysis. Columns list galaxy name, Galactic longitude, latitude, \ac{CMB}-frame redshift, host distance modulus inferred in the \Manticore\ analysis, and the predicted line-of-sight peculiar velocities from \Manticore\ and~\citetalias{Carrick_2015}. Uncertainties are $1\sigma$. All values assume \ac{SN} magnitude selection.}
    \label{tab:cepheid_hosts}
\end{table*}

\bsp
\label{lastpage}
\end{document}